\documentclass[prl,superscriptaddress,twocolumn,10pt,nofootinbib,aps, floatfix]{revtex4-1}
\usepackage{amssymb}
\usepackage{amsmath}
\usepackage{appendix}
\usepackage{times}
\usepackage{graphicx}
\usepackage{subeqnarray}

\usepackage{color}

\begin{document}

\title {Higher-Order Topological Phase in the Two-Dimensional Type-IV Magnet MgCr$_2$O$_4$}
	
\author{Xiaorong Zou}
\affiliation{School of Science, Qingdao University of Technology, Qingdao 266520, China}
\affiliation{Center for 2D Quantum Heterostructures (2DQH), Institute for Basic Science (IBS), Sungkyunkwan University, Seobu-ro 2066, Suwon 16419, Republic of Korea}

\author{Hyeon Suk Shin}
\affiliation{Center for 2D Quantum Heterostructures (2DQH), Institute for Basic Science (IBS), Sungkyunkwan University, Seobu-ro 2066, Suwon 16419, Republic of Korea}
\affiliation{Department of Chemistry, Sungkyunkwan University, Seobu-ro 2066, Suwon 16419, Republic of Korea}
\affiliation{Department of Energy, Sungkyunkwan University, Seobu-ro 2066, Suwon 16419, Republic of Korea}

\author{Yanmei Zang}
\affiliation{Department of Energy, Sungkyunkwan University, Seobu-ro 2066, Suwon 16419, Republic of Korea}
\affiliation{Department of Energy Science, Sungkyunkwan University, Seobu-ro 2066, Suwon 16419, Republic of Korea}

\author{Ying Dai}
\email{daiy60@sdu.edu.cn}
\affiliation {School of Physics, State Key Laboratory of Crystal Materials, Shandong University, Jinan 250100, China}

\author{Chengwang Niu}
\email{c.niu@sdu.edu.cn}
\affiliation {School of Physics, State Key Laboratory of Crystal Materials, Shandong University, Jinan 250100, China}

\author{Chang-Jong Kang}
\email{cjkang87@cnu.ac.kr}
\affiliation{Department of Physics, Chungnam National University, Daejeon 34134, Republic of Korea}
\affiliation{Institute for Sciences of the Universe, Chungnam National University, Daejeon 34134, Republic of Korea}

\author{Chang Woo Myung}
\email{cwmyung@skku.edu}
\affiliation{Center for 2D Quantum Heterostructures (2DQH), Institute for Basic Science (IBS), Sungkyunkwan University, Seobu-ro 2066, Suwon 16419, Republic of Korea}
\affiliation{Department of Energy, Sungkyunkwan University, Seobu-ro 2066, Suwon 16419, Republic of Korea}
\affiliation{Department of Energy Science, Sungkyunkwan University, Seobu-ro 2066, Suwon 16419, Republic of Korea}
\affiliation{Department of Quantum Information Engineering, Sungkyunkwan University, Seobu-ro 2066, Suwon, 16419, Korea}
	
\begin{abstract}
Type-IV two-dimensional (2D) magnetism--a newly classified collinear magnetic phase featuring nonrelativistic spin degeneracy and spin-orbit-coupling-induced momentum-dependent spin splitting--extends the symmetry classification of collinear magnets, opening new opportunities for unconventional topological quantum states. Here, we reveal that the recently proposed two-dimensional type-IV 2D magnet MgCr$_2$O$_4$ hosts an intrinsic higher-order topological insulating phase, featuring $\mathcal{C}_{3z}$-protected corner states and a nontrivial rotational topological invariant of $\chi^{(3)}$ = $\{-2,4\}$ with a quantized fractional corner charge of $4e/3$. Spin-orbit coupling breaks the spin-degeneracy-enforcing symmetry $[C_{2}||M_z]$ while preserving the crystalline $\mathcal{C}_{3z}$ rotational symmetry that protects the higher-order topological phase, thereby enabling  spin splitting to coexist with the nontrivial topology. Furthermore, the higher-order topological phase remains intact throughout a wide range of biaxial strains without band-gap closing and topological phase transition, demonstrating the robustness of the symmetry-protected topological state against external perturbations. Our work establishes a direct connection between type-IV magnetic system and higher-order topology, providing a new route for symmetry-engineered magnetic topological quantum states.	
\end{abstract}
	
\maketitle
\date{\today}

The interplay between magnetism and band topology has rapidly emerged as one of the central topics in condensed matter physics, providing a fertile platform for realizing exotic quantum phases with novel transport and spin-dependent functionalities~\cite{AFM2,AFM3,chang2020,2022natureAFM,2025NRMAFM,2025AFMNP,2026jacstopo}. Recent advances have uncovered a diverse range of  magnetic topological phenomena~\cite{2021NRPn12,2021Materreview,2023AFMSB,2025MTISA,2026MTIAFM}, including quantum anomalous Hall effect~\cite{2013scienceqah,2019qahprlxu,2025PRLQAH,2025PRLQAH2,2025PRLQAH3}, topological semimetals~\cite{2019MSMNPJ,2022PRLMSM,2024TSMPRA,2026arxivmsm} and the newly identified topological responses in altermagnets~\cite{2025PRLsuper,2025NRPAM,2026AFMAM,2026PRBAM,2026PRLAMODD}. These developments have demonstrated that the realization of topological quantum states is fundamentally governed by the underlying symmetry~\cite{database_mag,2021ARCTQC,2021NCTQC,2022NRMsymmetry,2023AMsymmetry,2024floNL,2026AFMtopo}, while each newly identified symmetry setting offers new possibilities for realizing previously unexplored topological phases. Accordingly, type-IV magnetism extends the symmetry classification of collinear magnetic phases beyond conventional ferromagnetism (type-I), antiferromagnetism (type-II), and altermagnetism (type-III)~\cite{2025PRLtype4}. While altermagnets exhibit spin splitting even without spin-orbit coupling (SOC), and conventional antiferromagnets maintain spin degeneracy regardless of SOC, type-IV magnets are uniquely governed by their spin-group symmetry: they exhibit symmetry-enforced spin degeneracy in the absence of SOC, and strictly require its inclusion to lift the degeneracy and develop momentum-dependent spin splitting~\cite{2026prltype4}. This distinctive symmetry setting naturally motivates the exploration of symmetry-protected topological phases in type-IV magnetic systems. However, the topological consequences of such unconventional symmetry constraints remain largely unexplored.

Among the various symmetry-protected topological phases, higher-order topological insulators (HOTIs) have generalized the conventional bulk-boundary correspondence~\cite{highorderfirst,2021NRPn1,2026PRLHOTI}. Unlike first-order topological insulators, which host gapless states on boundaries with one lower dimension~\cite{Hasan2010,2016TIRMP,2021cmsreview1}, HOTIs are characterized by topological boundary states localized on boundaries with two or more reduced dimensions~\cite{Schindler2018,highorderrenyafei,hoti2d}, giving rise to robust corner or hinge states protected by crystalline symmetries~\cite{highorderwyckoff,Hsu13255,2024PRLhotisymme}. Building upon this conceptual framework, higher-order topology has rapidly expanded beyond crystalline electronic systems to photonic systems~\cite{2019PRLHOTIPHOTON,2020NPHOTI,2025NChoti}, acoustic metamaterials~\cite{2018NMhotiacou,2019PRLhotiacou}, and topolectrical circuits~\cite{Imhof2018,Peterson2018,2023PRAHOTIcir}. More recently, the interplay between magnetism and higher-order topology has considerably enriched the HOTI family by broadening the symmetry landscape capable of hosting magnetic higher-order topological phases~\cite{higherordereuln2as2,2022npjfese,2024NLohe,2024PRBAMHOTI,2026ASAM}. Despite these advances, higher-order topology has so far been investigated predominantly in conventional magnetic systems. Whether the distinct symmetry framework of type-IV magnetism can stabilize higher-order topological phases therefore remains an important open question.

In this work, we investigate the higher-order topology of the two-dimensional type-IV magnet MgCr$_2$O$_4$ through first-principles calculations together with symmetry and topological analyses. Its intrinsic higher-order topological character can be confirmed through the bulk electronic structure, floating edge states, and symmetry-protected corner states in finite hexagonal nanoflakes. And we further uncover the symmetry mechanism underlying the higher-order topological phase by tracing the symmetry evolution induced by spin-orbit coupling, demonstrating that the crystalline $C_{3z}$ rotational symmetry continues to protect the higher-order topology despite the lifting of the spin-degeneracy-enforcing symmetry $[C_2||M_z]$. This symmetry mechanism is further corroborated by the nontrivial rotational topological invariant $\chi^{(3)}=(-2,\,4)$ and the corresponding quantized fractional corner charge $Q_{\rm corner}=4e/3$, providing complementary topological signatures of the higher-order topological phase. Finally, we systematically investigate the strain dependence of the higher-order topological phase and demonstrate that the symmetry-protected corner states remain well preserved over a wide range of biaxial strains (from $-5\%$ to $5\%$), highlighting the intrinsic robustness of the underlying topology. This work identifies higher-order topology as a fundamental manifestation of type-IV magnetic symmetry, thereby broadening the topological implications of type-IV magnets.

\begin{figure} 
	\centering
	\includegraphics[width=1\linewidth]{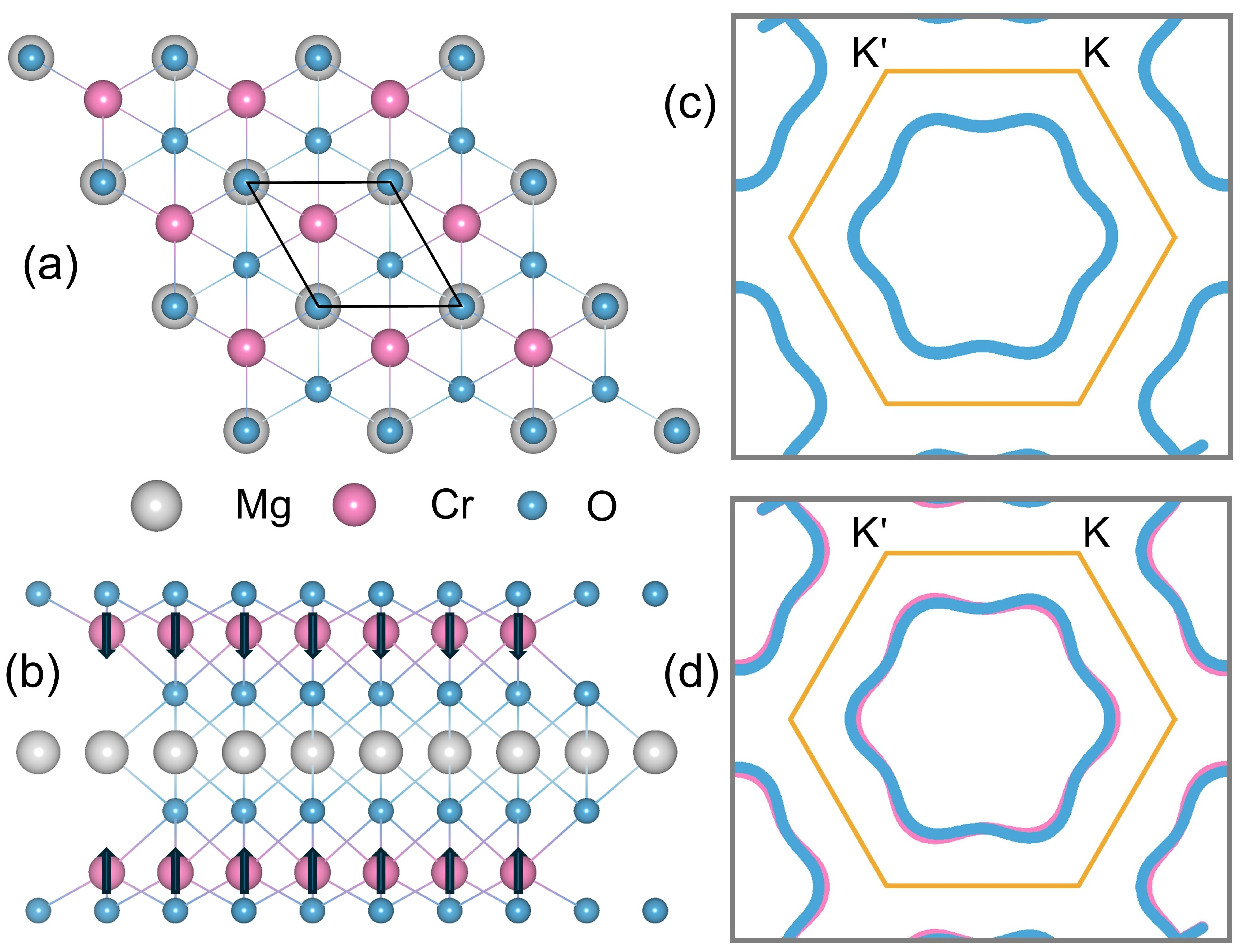}
	\caption{(a) Top and (b) side views of the crystal structure of the two-dimensional type-IV magnet MgCr$_2$O$_4$. The primitive unit cell is outlined in green, and the black arrows denote the magnetic moments of the Cr atoms. Constant-energy maps at $E-E_{\rm F}=-0.6$~eV calculated (c) without SOC and (d) with SOC.}
	\label{structure}
\end{figure}

First-principles calculations were performed within the framework of density functional theory (DFT) using the Vienna \textit{Ab Initio} Simulation Package (VASP)~\cite{Kresse,Kresse1}. The projector augmented-wave (PAW) method was employed together with the Perdew--Burke--Ernzerhof (PBE) exchange--correlation functional within the generalized gradient approximation (GGA)~\cite{pbevasp}. A plane-wave cutoff energy of 550~eV and a $\Gamma$-centered $9 \times 9 \times 1$ Monkhorst--Pack $k$-point mesh were adopted for Brillouin-zone sampling. To eliminate interactions between periodically repeated slabs, a vacuum layer of 20~\AA{} was introduced along the out-of-plane direction. Both the lattice parameters and atomic coordinates were fully optimized until the Hellmann--Feynman forces on each atom were below $0.01~\mathrm{eV/\AA}$, with the total-energy convergence criterion set to $10^{-6}$~eV. The on-site Coulomb interaction of the Cr-$3d$ electrons was treated within the Dudarev GGA + $U$ scheme using an effective Hubbard parameter of $U$ = 3 eV. Maximally localized Wannier functions were constructed using the WANNIER90 package~\cite{wannier90}, from which the Wannier tight-binding Hamiltonian was generated. The edge spectra and topological properties were subsequently calculated using the WannierTools package~\cite{WU2017}.

\begin{figure}
	\centering
	\includegraphics[width=1\linewidth]{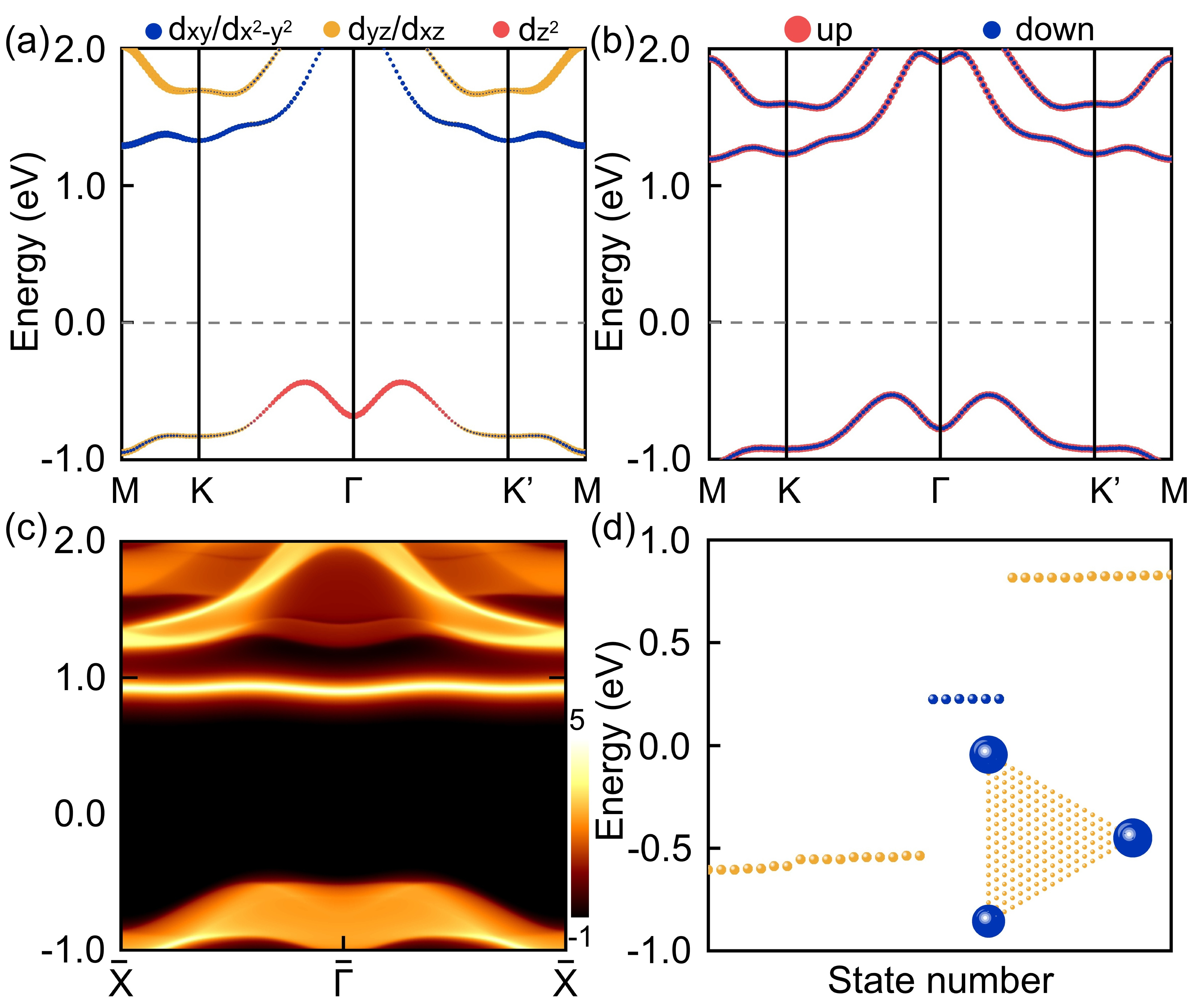}
	\caption{(a) Orbital-resolved band structure of the MgCr$_2$O$_4$ monolayer. (b) Spin-resolved band structure of the MgCr$_2$O$_4$ monolayer, showing spin degeneracy enforced by $\{C_{2}||M_z\}$. (c) Edge states and (d) discrete energy spectra of a triangular nanoflake for the 2D type-IV magnet MgCr$_2$O$_4$ monolayer. The Fermi level $E_F$ is set to be zero. The blue dots indicate the in-gap corner states. Insets show the corresponding real-space charge distributions of the six degenerate corner states, as marked by blue dots.} 
	\label{scf}
\end{figure}

The optimized crystal structure of monolayer MgCr$_2$O$_4$ is shown in Figs.~\ref{structure}(a) and (b). The primitive unit cell crystallizes in a hexagonal lattice with two symmetry-equivalent Cr atoms forming the magnetic sublattice. Total-energy calculations identify the antiparallel spin configuration as the magnetic ground state, lying 0.17~meV per unit cell below the parallel configuration. Within the recently developed spin-group formalism, the system belongs to the collinear spin layer group (cSLG) No.~187.1.2.2.L.1, satisfying the symmetry requirements of type-IV magnetism in the absence of SOC.  The spin-space symmetry $\{C_2||M_z\}$ enforces the relation $E(s,\mathbf{k})=E(-s,\mathbf{k})$ throughout the Brillouin zone, where $\mathbf{k} = (k_x, k_y)$ is the 2D momentum. Consistent with this symmetry constraint, the calculated electronic band structure shown in Fig.~\ref{scf}(b) exhibits complete spin degeneracy. And the system is an intrinsic semiconductor with an indirect band gap of 1.72 eV, with both the valence-band maximum and conduction-band minimum predominantly originating from Cr-$3d$ orbitals, as illustrated in Fig.~\ref{scf}(a).

\begin{figure}
	\centering
	\includegraphics[width=1\linewidth]{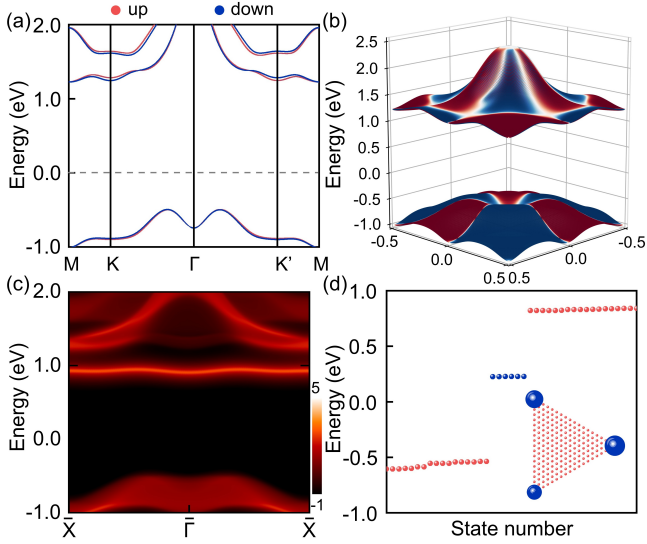}
	\caption{
		(a) Electronic band structure of the two-dimensional type-IV magnet MgCr$_2$O$_4$ with SOC, exhibiting momentum-dependent spin splitting. (b) Three-dimensional band structure of MgCr$_2$O$_4$ monolayer. (c) Edge states and (d) discrete energy spectra of a triangular nanoflake for the 2D type-IV magnet MgCr$_2$O$_4$ monolayer with SOC. The Fermi level $E_F$ is set to be zero. The blue dots indicate the in-gap corner states. Insets show the corresponding real-space charge distributions of the six degenerate corner states, as marked by blue dots.}
	\label{soc}
\end{figure}

To identify the topological nature of monolayer MgCr$_2$O$_4$, we first examine its generalized bulk-boundary correspondence. Since first- and second-order topological phases exhibit distinct boundary electronic structures, the edge spectrum of a semi-infinite system provides a direct criterion for distinguishing between them. We therefore calculate the edge spectrum of a semi-infinite system using the iterative Green's-function method based on the Wannier tight-binding Hamiltonian. As shown in Fig.~\ref{scf}(c), no gapless edge mode traverses the bulk band gap. Instead, isolated floating edge states emerge inside the insulating gap, indicating that the nontrivial topology is manifested through lower-dimensional boundary states, as expected for a higher-order topological phase. 

A natural manifestation of the underlying topology is provided by a finite triangular nanoflake preserving the crystalline $C_{3z}$ rotational symmetry. The corresponding energy spectrum, depicted in Fig.~\ref{scf}(d), exhibits six nearly degenerate in-gap states that are well separated from the bulk spectrum. The real-space distributions of these in-gap states are presented in Fig.~\ref{scf}(d), where the wave functions are strongly localized at the three corners of the nanoflake with negligible weight distributed in the bulk and along the edges. The coexistence of floating edge states in the semi-infinite geometry and corner-localized in-gap states in the finite triangular nanoflake establishes the generalized bulk-boundary correspondence characteristic of a second-order topological insulator, thereby identifying monolayer MgCr$_2$O$_4$ as an intrinsic higher-order topological insulator.  Their robustness against different corner terminations is further confirmed in Fig. S1, providing additional evidence for their higher-order
topological origin.

To further characterize the higher-order topology from the symmetry perspective, we evaluate the rotational topological invariant associated with the preserved crystalline $C_{3z}$ symmetry~\cite{highorderinvariants,c3invariant,higher2020invariant}. The symmetry-resolved indices are obtained from the $C_{3z}$ rotation eigenvalues of the occupied bands and are defined as
\begin{equation}
	[K_{p}^{(3)}]=\#K_{p}^{(3)}-\#\Gamma_{p}^{(3)},
\end{equation}
where $\#K_{p}^{(3)}$ and $\#\Gamma_{p}^{(3)}$ denote the numbers of occupied bands carrying the $p$th eigenvalue $e^{i\frac{2\pi}{3}(p-1)}$ ($p=1,2,3$) of the $C_{3z}$ rotation operator at the $K$ and $\Gamma$ points, respectively. The corresponding rotational topological invariant and fractional corner charge are expressed as
\begin{align}
	\chi^{(3)} &= \left([K_{1}^{(3)}],[K_{2}^{(3)}]\right),\\
	Q_{\rm corner}^{(3)} &= \frac{e}{3}[K_{2}^{(3)}]
	\quad\mathrm{mod}\ e.
\end{align}
In the absence of SOC, the spin-up and spin-down sectors remain completely decoupled, and the symmetry indicators are therefore evaluated within the spinless representation. The calculated rotational invariant is
\[
\chi^{(3)}=(-4,2),
\]
corresponding to a quantized fractional corner charge of
\[
Q_{\rm corner}^{(3)}=\frac{2e}{3}.
\]
The quantized fractional corner charge is fully consistent with the corner states observed in the finite triangular nanoflake, providing a symmetry-based characterization of the higher-order topological phase.

We next analyze the symmetry evolution induced by spin-orbit coupling (SOC) and its implications for the electronic structure and topology of monolayer MgCr$_2$O$_4$. Upon introducing SOC, the symmetry of the system evolves from the collinear spin layer group (cSLG) No.~187.1.2.2.L.1 to the corresponding magnetic layer group (MLG) No.~78.5.514. Consequently, the spin-space symmetry operation $\{C_2||M_z\}$, which enforces the relation $E(s,\mathbf{k})=E(-s,\mathbf{k})$ in the absence of SOC, is no longer preserved. In contrast, the crystalline threefold rotational symmetry $C_{3z}$, protecting the higher-order topology, is maintained. As shown in Fig.~\ref{soc}(a), the spin degeneracy is completely lifted along the high-symmetry path, giving rise to pronounced momentum-dependent spin splitting, further highlighted by the three-dimensional band structure in Fig.~\ref{soc}(b). Notably, the preservation of the crystalline $C_{3z}$ symmetry suggests the higher-order topological phase to remain robust despite the SOC-induced lifting of spin degeneracy.

Guided by this symmetry analysis, we investigate the boundary signatures of the higher-order topology after the inclusion of spin-orbit coupling. As shown in Figs.~3(c) and (d), the floating edge states remain inside the bulk gap, while the finite triangular nanoflake continues to host six nearly degenerate in-gap corner states localized at the three crystalline corners. The persistence of these characteristic boundary states demonstrates that the higher-order topological phase is preserved after the inclusion of SOC. Since the crystalline $C_{3z}$ rotational symmetry is retained, the higher-order topology can still be characterized by the associated symmetry-based rotational invariant. Different from the spinless case, the rotational invariant is evaluated within the double-valued representation in the presence of SOC, where the occupied bands are labeled by the $C_{3z}$ eigenvalues $e^{i\frac{\pi}{3}(2p-1)}$ ($p=1,2,3$). The calculated rotational invariant is
\begin{align}
	\chi^{(3)} &= \left([K_{1}^{(3)}],[K_{2}^{(3)}]\right)=(-2,4),
\end{align}
and the fractional corner charge is
\begin{align}
	Q_{\rm corner}^{(3)}
	&=\frac{2e}{3}\left([K_{1}^{(3)}]+[K_{2}^{(3)}]\right)
	\quad\mathrm{mod}\ 2e
	=\frac{4e}{3}.
\end{align}
Remarkably, the fractional corner charge obtained with SOC is exactly equal to the sum of the spin-resolved fractional corner charges in the absence of SOC, demonstrating the consistency of the symmetry-based topological characterization before and after spin-orbit coupling~\cite{2022PRBHOTI1}. Together with the nontrivial rotational invariant, these results provide coherent picture that the higher-order topological phase remains intact.

\begin{figure}
	\centering
	\includegraphics[width=1\linewidth]{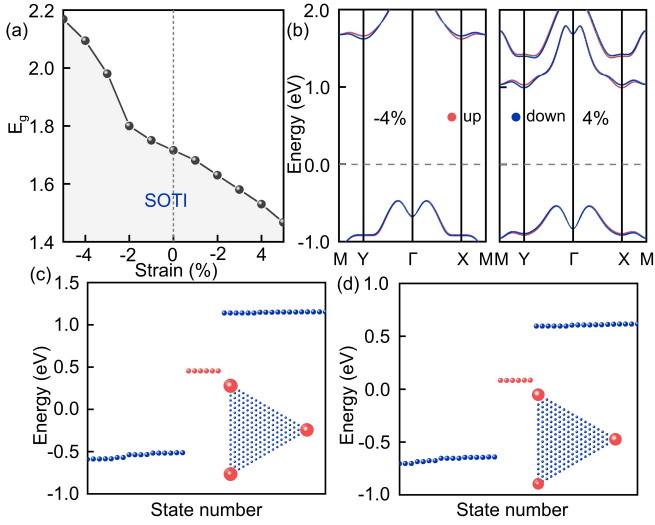}
	\caption{(a) Evolution of the bulk band gap of monolayer MgCr$_2$O$_4$ under biaxial strain, indicating that no band-gap closing occurs throughout the investigated strain range. (b) Electronic band structures under compressive ($\eta=-4\%$) and tensile ($\eta=4\%$) strains. Discrete energy spectra of finite triangular nanoflakes under (c) $\eta=-4\%$ and (d) $\eta=4\%$. The Fermi level $E_{\rm F}$ is set to zero. The red dots denote the in-gap corner states. Insets show the corresponding real-space wave-function distributions of the six degenerate corner states, as marked by red dots.} 
	\label{strain}
\end{figure}

Finally, we investigate the robustness of the higher-order topological phase against biaxial strain in monolayer MgCr$_2$O$_4$. The biaxial strain is defined as
\[
\eta=\frac{a-a_0}{a_0}\times100\%,
\]
where $a_0$ and $a$ denote the equilibrium and strained lattice constants, respectively. The biaxial strain is continuously varied from $-5\%$ to $5\%$, and the evolution of the bulk electronic structure is systematically tracked. Importantly, the type-IV magnetic state persists throughout the entire strain range considered (see Table S1). As shown in Fig. 4(a), the bulk band gap remains open and consistently exceeds 1.4 eV under biaxial strain. Consequently, no band-gap closing and reopening is observed over the investigated strain range, excluding the occurrence of a strain-induced topological phase transition.

To further verify the robustness of the boundary topology, two representative strained structures with $\eta=-4\%$ and $\eta=4\%$ are selected for detailed analysis. The corresponding finite-size spectra and real-space wave-function distributions are shown in Figs.~\ref{strain}(c) and (d). Under both compressive and tensile strains, six nearly degenerate in-gap corner states remain well separated from the bulk spectrum, while their wave functions continue to be strongly localized at the three corners of the triangular nanoflake. Collectively, the absence of strain-induced band-gap closing, the unchanged nontrivial rotational topological invariant $\chi^{(3)}=(-2,\,4)$, and the persistence of the characteristic corner states demonstrate that the higher-order topological phase remains robust throughout the investigated strain range.

In summary, we demonstrate that the MgCr$_2$O$_4$ monolayer realizes an intrinsic higher-order topological insulating phase in the type-IV magnetic state. The higher-order topology is fully preserved throughout the symmetry evolution from the spin group to the magnetic layer group, despite the symmetry reduction induced by spin-orbit coupling. The resulting topological phase is characterized by topological corner states, a nontrivial $\mathcal{C}_3$ rotational invariant, and quantized fractional corner charge, while remaining robust across the considered biaxial strain range. Taken together, these findings clearly identify the higher-order topology as an intrinsic and robust property of the type-IV magnetic phase in monolayer MgCr$_2$O$_4$. Our work extends the topological properties of type-IV magnets and provides a representative material platform for further investigating higher-order topology in this emerging class of magnetic materials.

This work was supported by Institute for Basic Science (IBS-R036-D1) and the Taishan Scholar Program of Shandong Province. We are grateful for the computational support from the Korea Institute of Science and Technology Information (KISTI) (KSC-2023-CRE-0355, KSC-2023-CRE-0261, KSC-2023-CRE-0502, KSC-2024-CRE-0144, KSC-2024-CRE-0088, KSC-2024-CRE-0117). Computational work for this research was partially performed on the Olaf supercomputer supported by IBS Research Solution Center and GPU cluster supported by Ministry of Science and ICT (MSIT) and the National IT Industry Promotion Agency (NIPA).


\begin{thebibliography}{69}%
	\makeatletter
	\providecommand \@ifxundefined [1]{%
		\@ifx{#1\undefined}
	}%
	\providecommand \@ifnum [1]{%
		\ifnum #1\expandafter \@firstoftwo
		\else \expandafter \@secondoftwo
		\fi
	}%
	\providecommand \@ifx [1]{%
		\ifx #1\expandafter \@firstoftwo
		\else \expandafter \@secondoftwo
		\fi
	}%
	\providecommand \natexlab [1]{#1}%
	\providecommand \enquote  [1]{``#1''}%
	\providecommand \bibnamefont  [1]{#1}%
	\providecommand \bibfnamefont [1]{#1}%
	\providecommand \citenamefont [1]{#1}%
	\providecommand \href@noop [0]{\@secondoftwo}%
	\providecommand \href [0]{\begingroup \@sanitize@url \@href}%
	\providecommand \@href[1]{\@@startlink{#1}\@@href}%
	\providecommand \@@href[1]{\endgroup#1\@@endlink}%
	\providecommand \@sanitize@url [0]{\catcode `\\12\catcode `\$12\catcode
		`\&12\catcode `\#12\catcode `\^12\catcode `\_12\catcode `\%12\relax}%
	\providecommand \@@startlink[1]{}%
	\providecommand \@@endlink[0]{}%
	\providecommand \url  [0]{\begingroup\@sanitize@url \@url }%
	\providecommand \@url [1]{\endgroup\@href {#1}{\urlprefix }}%
	\providecommand \urlprefix  [0]{URL }%
	\providecommand \Eprint [0]{\href }%
	\providecommand \doibase [0]{http://dx.doi.org/}%
	\providecommand \selectlanguage [0]{\@gobble}%
	\providecommand \bibinfo  [0]{\@secondoftwo}%
	\providecommand \bibfield  [0]{\@secondoftwo}%
	\providecommand \translation [1]{[#1]}%
	\providecommand \BibitemOpen [0]{}%
	\providecommand \bibitemStop [0]{}%
	\providecommand \bibitemNoStop [0]{.\EOS\space}%
	\providecommand \EOS [0]{\spacefactor3000\relax}%
	\providecommand \BibitemShut  [1]{\csname bibitem#1\endcsname}%
	\let\auto@bib@innerbib\@empty
	\bibitem [{\citenamefont {Jungwirth}\ \emph {et~al.}(2018)\citenamefont
		{Jungwirth}, \citenamefont {Sinova}, \citenamefont {Manchon}, \citenamefont
		{Marti}, \citenamefont {Wunderlich},\ and\ \citenamefont {Felser}}]{AFM2}%
	\BibitemOpen
	\bibfield  {author} {\bibinfo {author} {\bibfnamefont {T.}~\bibnamefont
			{Jungwirth}}, \bibinfo {author} {\bibfnamefont {J.}~\bibnamefont {Sinova}},
		\bibinfo {author} {\bibfnamefont {A.}~\bibnamefont {Manchon}}, \bibinfo
		{author} {\bibfnamefont {X.}~\bibnamefont {Marti}}, \bibinfo {author}
		{\bibfnamefont {J.}~\bibnamefont {Wunderlich}}, \ and\ \bibinfo {author}
		{\bibfnamefont {C.}~\bibnamefont {Felser}},\ }\href {\doibase
		10.1038/s41567-018-0063-6} {\bibfield  {journal} {\bibinfo  {journal} {Nat.
				Phys.}\ }\textbf {\bibinfo {volume} {14}},\ \bibinfo {pages} {200} (\bibinfo
		{year} {2018})}\BibitemShut {NoStop}%
	\bibitem [{\citenamefont {Šmejkal}\ \emph {et~al.}(2018)\citenamefont
		{Šmejkal}, \citenamefont {Mokrousov}, \citenamefont {Yan},\ and\
		\citenamefont {MacDonald}}]{AFM3}%
	\BibitemOpen
	\bibfield  {author} {\bibinfo {author} {\bibfnamefont {L.}~\bibnamefont
			{Šmejkal}}, \bibinfo {author} {\bibfnamefont {Y.}~\bibnamefont {Mokrousov}},
		\bibinfo {author} {\bibfnamefont {B.}~\bibnamefont {Yan}}, \ and\ \bibinfo
		{author} {\bibfnamefont {A.~H.}\ \bibnamefont {MacDonald}},\ }\href {\doibase
		10.1038/s41567-018-0064-5} {\bibfield  {journal} {\bibinfo  {journal} {Nat.
				Phys.}\ }\textbf {\bibinfo {volume} {14}},\ \bibinfo {pages} {242} (\bibinfo
		{year} {2018})}\BibitemShut {NoStop}%
	\bibitem [{\citenamefont {Chang}(2020)}]{chang2020}%
	\BibitemOpen
	\bibfield  {author} {\bibinfo {author} {\bibfnamefont {C.-Z.}\ \bibnamefont
			{Chang}},\ }\href@noop {} {\bibfield  {journal} {\bibinfo  {journal} {Nat.
				Mater.}\ }\textbf {\bibinfo {volume} {19}},\ \bibinfo {pages} {484} (\bibinfo
		{year} {2020})}\BibitemShut {NoStop}%
	\bibitem [{\citenamefont {Bernevig}\ \emph {et~al.}(2022)\citenamefont
		{Bernevig}, \citenamefont {Felser},\ and\ \citenamefont
		{Beidenkopf}}]{2022natureAFM}%
	\BibitemOpen
	\bibfield  {author} {\bibinfo {author} {\bibfnamefont {B.~A.}\ \bibnamefont
			{Bernevig}}, \bibinfo {author} {\bibfnamefont {C.}~\bibnamefont {Felser}}, \
		and\ \bibinfo {author} {\bibfnamefont {H.}~\bibnamefont {Beidenkopf}},\
	}\href {\doibase 10.1038/s41586-021-04105-x} {\bibfield  {journal} {\bibinfo
			{journal} {Nature}\ }\textbf {\bibinfo {volume} {603}},\ \bibinfo {pages}
		{41} (\bibinfo {year} {2022})}\BibitemShut {NoStop}%
	\bibitem [{\citenamefont {Rimmler}\ \emph {et~al.}(2025)\citenamefont
		{Rimmler}, \citenamefont {Pal},\ and\ \citenamefont {Parkin}}]{2025NRMAFM}%
	\BibitemOpen
	\bibfield  {author} {\bibinfo {author} {\bibfnamefont {B.~H.}\ \bibnamefont
			{Rimmler}}, \bibinfo {author} {\bibfnamefont {B.}~\bibnamefont {Pal}}, \ and\
		\bibinfo {author} {\bibfnamefont {S.~S.~P.}\ \bibnamefont {Parkin}},\ }\href
	{\doibase 10.1038/s41578-024-00706-w} {\bibfield  {journal} {\bibinfo
			{journal} {Nat. Rev. Mater.}\ }\textbf {\bibinfo {volume} {10}},\ \bibinfo
		{pages} {109} (\bibinfo {year} {2025})}\BibitemShut {NoStop}%
	\bibitem [{\citenamefont {Bielinski}\ \emph {et~al.}(2025)\citenamefont
		{Bielinski}, \citenamefont {Chari}, \citenamefont {May-Mann}, \citenamefont
		{Kim}, \citenamefont {Zwettler}, \citenamefont {Deng}, \citenamefont
		{Aishwarya}, \citenamefont {Roychowdhury}, \citenamefont {Shekhar},
		\citenamefont {Hashimoto}, \citenamefont {Lu}, \citenamefont {Yan},
		\citenamefont {Felser}, \citenamefont {Madhavan}, \citenamefont {Shen},
		\citenamefont {Hughes},\ and\ \citenamefont {Mahmood}}]{2025AFMNP}%
	\BibitemOpen
	\bibfield  {author} {\bibinfo {author} {\bibfnamefont {N.}~\bibnamefont
			{Bielinski}}, \bibinfo {author} {\bibfnamefont {R.}~\bibnamefont {Chari}},
		\bibinfo {author} {\bibfnamefont {J.}~\bibnamefont {May-Mann}}, \bibinfo
		{author} {\bibfnamefont {S.}~\bibnamefont {Kim}}, \bibinfo {author}
		{\bibfnamefont {J.}~\bibnamefont {Zwettler}}, \bibinfo {author}
		{\bibfnamefont {Y.}~\bibnamefont {Deng}}, \bibinfo {author} {\bibfnamefont
			{A.}~\bibnamefont {Aishwarya}}, \bibinfo {author} {\bibfnamefont
			{S.}~\bibnamefont {Roychowdhury}}, \bibinfo {author} {\bibfnamefont
			{C.}~\bibnamefont {Shekhar}}, \bibinfo {author} {\bibfnamefont
			{M.}~\bibnamefont {Hashimoto}}, \bibinfo {author} {\bibfnamefont
			{D.}~\bibnamefont {Lu}}, \bibinfo {author} {\bibfnamefont {J.}~\bibnamefont
			{Yan}}, \bibinfo {author} {\bibfnamefont {C.}~\bibnamefont {Felser}},
		\bibinfo {author} {\bibfnamefont {V.}~\bibnamefont {Madhavan}}, \bibinfo
		{author} {\bibfnamefont {Z.-X.}\ \bibnamefont {Shen}}, \bibinfo {author}
		{\bibfnamefont {T.~L.}\ \bibnamefont {Hughes}}, \ and\ \bibinfo {author}
		{\bibfnamefont {F.}~\bibnamefont {Mahmood}},\ }\href {\doibase
		10.1038/s41567-024-02769-6} {\bibfield  {journal} {\bibinfo  {journal} {Nat.
				Phys.}\ }\textbf {\bibinfo {volume} {21}},\ \bibinfo {pages} {458} (\bibinfo
		{year} {2025})}\BibitemShut {NoStop}%
	\bibitem [{\citenamefont {Yu}\ and\ \citenamefont
		{Heine}(2026)}]{2026jacstopo}%
	\BibitemOpen
	\bibfield  {author} {\bibinfo {author} {\bibfnamefont {H.}~\bibnamefont
			{Yu}}\ and\ \bibinfo {author} {\bibfnamefont {T.}~\bibnamefont {Heine}},\
	}\href {\doibase 10.1021/jacs.5c21206} {\bibfield  {journal} {\bibinfo
			{journal} {J. Am. Chem. Soc.}\ }\textbf {\bibinfo {volume} {148}},\ \bibinfo
		{pages} {13822} (\bibinfo {year} {2026})}\BibitemShut {NoStop}%
	\bibitem [{\citenamefont {Xiao}\ and\ \citenamefont {Yan}(2021)}]{2021NRPn12}%
	\BibitemOpen
	\bibfield  {author} {\bibinfo {author} {\bibfnamefont {J.}~\bibnamefont
			{Xiao}}\ and\ \bibinfo {author} {\bibfnamefont {B.}~\bibnamefont {Yan}},\
	}\href {\doibase 10.1038/s42254-021-00292-8} {\bibfield  {journal} {\bibinfo
			{journal} {Nat. Rev. Phys.}\ }\textbf {\bibinfo {volume} {3}},\ \bibinfo
		{pages} {283} (\bibinfo {year} {2021})}\BibitemShut {NoStop}%
	\bibitem [{\citenamefont {Polash}\ \emph {et~al.}(2021)\citenamefont {Polash},
		\citenamefont {Yalameha}, \citenamefont {Zhou}, \citenamefont {Ahadi},
		\citenamefont {Nourbakhsh},\ and\ \citenamefont {Vashaee}}]{2021Materreview}%
	\BibitemOpen
	\bibfield  {author} {\bibinfo {author} {\bibfnamefont {M.~M.~H.}\
			\bibnamefont {Polash}}, \bibinfo {author} {\bibfnamefont {S.}~\bibnamefont
			{Yalameha}}, \bibinfo {author} {\bibfnamefont {H.}~\bibnamefont {Zhou}},
		\bibinfo {author} {\bibfnamefont {K.}~\bibnamefont {Ahadi}}, \bibinfo
		{author} {\bibfnamefont {Z.}~\bibnamefont {Nourbakhsh}}, \ and\ \bibinfo
		{author} {\bibfnamefont {D.}~\bibnamefont {Vashaee}},\ }\href {\doibase
		https://doi.org/10.1016/j.mser.2021.100620} {\bibfield  {journal} {\bibinfo
			{journal} {Mater. Sci. Eng. R Rep.}\ }\textbf {\bibinfo {volume} {145}},\
		\bibinfo {pages} {100620} (\bibinfo {year} {2021})}\BibitemShut {NoStop}%
	\bibitem [{\citenamefont {Zhang}\ \emph {et~al.}(2023)\citenamefont {Zhang},
		\citenamefont {Wang}, \citenamefont {He}, \citenamefont {Wang}, \citenamefont
		{Yu}, \citenamefont {Liu}, \citenamefont {Liu},\ and\ \citenamefont
		{Cheng}}]{2023AFMSB}%
	\BibitemOpen
	\bibfield  {author} {\bibinfo {author} {\bibfnamefont {X.}~\bibnamefont
			{Zhang}}, \bibinfo {author} {\bibfnamefont {X.}~\bibnamefont {Wang}},
		\bibinfo {author} {\bibfnamefont {T.}~\bibnamefont {He}}, \bibinfo {author}
		{\bibfnamefont {L.}~\bibnamefont {Wang}}, \bibinfo {author} {\bibfnamefont
			{W.-W.}\ \bibnamefont {Yu}}, \bibinfo {author} {\bibfnamefont
			{Y.}~\bibnamefont {Liu}}, \bibinfo {author} {\bibfnamefont {G.}~\bibnamefont
			{Liu}}, \ and\ \bibinfo {author} {\bibfnamefont {Z.}~\bibnamefont {Cheng}},\
	}\href {\doibase https://doi.org/10.1016/j.scib.2023.09.004} {\bibfield
		{journal} {\bibinfo  {journal} {Sci. Bull.}\ }\textbf {\bibinfo {volume}
			{68}},\ \bibinfo {pages} {2639} (\bibinfo {year} {2023})}\BibitemShut
	{NoStop}%
	\bibitem [{\citenamefont {Robredo}\ \emph {et~al.}(2025)\citenamefont
		{Robredo}, \citenamefont {Xu}, \citenamefont {Jiang}, \citenamefont {Felser},
		\citenamefont {Bernevig}, \citenamefont {Elcoro}, \citenamefont {Regnault},\
		and\ \citenamefont {Vergniory}}]{2025MTISA}%
	\BibitemOpen
	\bibfield  {author} {\bibinfo {author} {\bibfnamefont {I.}~\bibnamefont
			{Robredo}}, \bibinfo {author} {\bibfnamefont {Y.}~\bibnamefont {Xu}},
		\bibinfo {author} {\bibfnamefont {Y.}~\bibnamefont {Jiang}}, \bibinfo
		{author} {\bibfnamefont {C.}~\bibnamefont {Felser}}, \bibinfo {author}
		{\bibfnamefont {B.~A.}\ \bibnamefont {Bernevig}}, \bibinfo {author}
		{\bibfnamefont {L.}~\bibnamefont {Elcoro}}, \bibinfo {author} {\bibfnamefont
			{N.}~\bibnamefont {Regnault}}, \ and\ \bibinfo {author} {\bibfnamefont
			{M.~G.}\ \bibnamefont {Vergniory}},\ }\href {\doibase 10.1126/sciadv.adv8780}
	{\bibfield  {journal} {\bibinfo  {journal} {Sci. Adv.}\ }\textbf {\bibinfo
			{volume} {11}},\ \bibinfo {pages} {eadv8780} (\bibinfo {year}
		{2025})}\BibitemShut {NoStop}%
	\bibitem [{\citenamefont {Dixit}\ \emph {et~al.}(2026)\citenamefont {Dixit},
		\citenamefont {Yang}, \citenamefont {Li}, \citenamefont {Chen}, \citenamefont
		{Jia},\ and\ \citenamefont {Wang}}]{2026MTIAFM}%
	\BibitemOpen
	\bibfield  {author} {\bibinfo {author} {\bibfnamefont {B.}~\bibnamefont
			{Dixit}}, \bibinfo {author} {\bibfnamefont {Y.}~\bibnamefont {Yang}},
		\bibinfo {author} {\bibfnamefont {S.}~\bibnamefont {Li}}, \bibinfo {author}
		{\bibfnamefont {Y.-C.}\ \bibnamefont {Chen}}, \bibinfo {author}
		{\bibfnamefont {Q.}~\bibnamefont {Jia}}, \ and\ \bibinfo {author}
		{\bibfnamefont {J.-P.}\ \bibnamefont {Wang}},\ }\href {\doibase
		https://doi.org/10.1002/adfm.202528026} {\bibfield  {journal} {\bibinfo
			{journal} {Adv. Funct. Mater.}\ }\textbf {\bibinfo {volume} {36}},\ \bibinfo
		{pages} {e28026} (\bibinfo {year} {2026})}\BibitemShut {NoStop}%
	\bibitem [{\citenamefont {Chang}\ \emph {et~al.}(2013)\citenamefont {Chang},
		\citenamefont {Zhang}, \citenamefont {Feng}, \citenamefont {Shen},
		\citenamefont {Zhang}, \citenamefont {Guo}, \citenamefont {Li}, \citenamefont
		{Ou}, \citenamefont {Wei}, \citenamefont {Wang}, \citenamefont {Ji},
		\citenamefont {Feng}, \citenamefont {Ji}, \citenamefont {Chen}, \citenamefont
		{Jia}, \citenamefont {Dai}, \citenamefont {Fang}, \citenamefont {Zhang},
		\citenamefont {He}, \citenamefont {Wang}, \citenamefont {Lu}, \citenamefont
		{Ma},\ and\ \citenamefont {Xue}}]{2013scienceqah}%
	\BibitemOpen
	\bibfield  {author} {\bibinfo {author} {\bibfnamefont {C.-Z.}\ \bibnamefont
			{Chang}}, \bibinfo {author} {\bibfnamefont {J.}~\bibnamefont {Zhang}},
		\bibinfo {author} {\bibfnamefont {X.}~\bibnamefont {Feng}}, \bibinfo {author}
		{\bibfnamefont {J.}~\bibnamefont {Shen}}, \bibinfo {author} {\bibfnamefont
			{Z.}~\bibnamefont {Zhang}}, \bibinfo {author} {\bibfnamefont
			{M.}~\bibnamefont {Guo}}, \bibinfo {author} {\bibfnamefont {K.}~\bibnamefont
			{Li}}, \bibinfo {author} {\bibfnamefont {Y.}~\bibnamefont {Ou}}, \bibinfo
		{author} {\bibfnamefont {P.}~\bibnamefont {Wei}}, \bibinfo {author}
		{\bibfnamefont {L.-L.}\ \bibnamefont {Wang}}, \bibinfo {author}
		{\bibfnamefont {Z.-Q.}\ \bibnamefont {Ji}}, \bibinfo {author} {\bibfnamefont
			{Y.}~\bibnamefont {Feng}}, \bibinfo {author} {\bibfnamefont {S.}~\bibnamefont
			{Ji}}, \bibinfo {author} {\bibfnamefont {X.}~\bibnamefont {Chen}}, \bibinfo
		{author} {\bibfnamefont {J.}~\bibnamefont {Jia}}, \bibinfo {author}
		{\bibfnamefont {X.}~\bibnamefont {Dai}}, \bibinfo {author} {\bibfnamefont
			{Z.}~\bibnamefont {Fang}}, \bibinfo {author} {\bibfnamefont {S.-C.}\
			\bibnamefont {Zhang}}, \bibinfo {author} {\bibfnamefont {K.}~\bibnamefont
			{He}}, \bibinfo {author} {\bibfnamefont {Y.}~\bibnamefont {Wang}}, \bibinfo
		{author} {\bibfnamefont {L.}~\bibnamefont {Lu}}, \bibinfo {author}
		{\bibfnamefont {X.-C.}\ \bibnamefont {Ma}}, \ and\ \bibinfo {author}
		{\bibfnamefont {Q.-K.}\ \bibnamefont {Xue}},\ }\href {\doibase
		doi:10.1126/science.1234414} {\bibfield  {journal} {\bibinfo  {journal}
			{Science}\ }\textbf {\bibinfo {volume} {340}},\ \bibinfo {pages} {167}
		(\bibinfo {year} {2013})}\BibitemShut {NoStop}%
	\bibitem [{\citenamefont {Sun}\ \emph {et~al.}(2019)\citenamefont {Sun},
		\citenamefont {Xia}, \citenamefont {Chen}, \citenamefont {Zhang},
		\citenamefont {Liu}, \citenamefont {Yao}, \citenamefont {Tang}, \citenamefont
		{Zhao}, \citenamefont {Xu},\ and\ \citenamefont {Liu}}]{2019qahprlxu}%
	\BibitemOpen
	\bibfield  {author} {\bibinfo {author} {\bibfnamefont {H.}~\bibnamefont
			{Sun}}, \bibinfo {author} {\bibfnamefont {B.}~\bibnamefont {Xia}}, \bibinfo
		{author} {\bibfnamefont {Z.}~\bibnamefont {Chen}}, \bibinfo {author}
		{\bibfnamefont {Y.}~\bibnamefont {Zhang}}, \bibinfo {author} {\bibfnamefont
			{P.}~\bibnamefont {Liu}}, \bibinfo {author} {\bibfnamefont {Q.}~\bibnamefont
			{Yao}}, \bibinfo {author} {\bibfnamefont {H.}~\bibnamefont {Tang}}, \bibinfo
		{author} {\bibfnamefont {Y.}~\bibnamefont {Zhao}}, \bibinfo {author}
		{\bibfnamefont {H.}~\bibnamefont {Xu}}, \ and\ \bibinfo {author}
		{\bibfnamefont {Q.}~\bibnamefont {Liu}},\ }\href {\doibase
		10.1103/PhysRevLett.123.096401} {\bibfield  {journal} {\bibinfo  {journal}
			{Phys. Rev. Lett.}\ }\textbf {\bibinfo {volume} {123}},\ \bibinfo {pages}
		{096401} (\bibinfo {year} {2019})}\BibitemShut {NoStop}%
	\bibitem [{\citenamefont {Liang}\ \emph {et~al.}(2025)\citenamefont {Liang},
		\citenamefont {Li}, \citenamefont {An}, \citenamefont {Ren}, \citenamefont
		{Qiao},\ and\ \citenamefont {Niu}}]{2025PRLQAH}%
	\BibitemOpen
	\bibfield  {author} {\bibinfo {author} {\bibfnamefont {W.}~\bibnamefont
			{Liang}}, \bibinfo {author} {\bibfnamefont {Z.}~\bibnamefont {Li}}, \bibinfo
		{author} {\bibfnamefont {J.}~\bibnamefont {An}}, \bibinfo {author}
		{\bibfnamefont {Y.}~\bibnamefont {Ren}}, \bibinfo {author} {\bibfnamefont
			{Z.}~\bibnamefont {Qiao}}, \ and\ \bibinfo {author} {\bibfnamefont
			{Q.}~\bibnamefont {Niu}},\ }\href {\doibase 10.1103/PhysRevLett.134.116603}
	{\bibfield  {journal} {\bibinfo  {journal} {Phys. Rev. Lett.}\ }\textbf
		{\bibinfo {volume} {134}},\ \bibinfo {pages} {116603} (\bibinfo {year}
		{2025})}\BibitemShut {NoStop}%
	\bibitem [{\citenamefont {Wan}\ \emph {et~al.}(2025)\citenamefont {Wan},
		\citenamefont {Liu},\ and\ \citenamefont {Sun}}]{2025PRLQAH2}%
	\BibitemOpen
	\bibfield  {author} {\bibinfo {author} {\bibfnamefont {Y.-H.}\ \bibnamefont
			{Wan}}, \bibinfo {author} {\bibfnamefont {P.-Y.}\ \bibnamefont {Liu}}, \ and\
		\bibinfo {author} {\bibfnamefont {Q.-F.}\ \bibnamefont {Sun}},\ }\href
	{\doibase 10.1103/8vs2-jvc4} {\bibfield  {journal} {\bibinfo  {journal}
			{Phys. Rev. Lett.}\ }\textbf {\bibinfo {volume} {135}},\ \bibinfo {pages}
		{186302} (\bibinfo {year} {2025})}\BibitemShut {NoStop}%
	\bibitem [{\citenamefont {Dai}\ \emph {et~al.}(2025)\citenamefont {Dai},
		\citenamefont {Zhu},\ and\ \citenamefont {He}}]{2025PRLQAH3}%
	\BibitemOpen
	\bibfield  {author} {\bibinfo {author} {\bibfnamefont {Z.}~\bibnamefont
			{Dai}}, \bibinfo {author} {\bibfnamefont {X.}~\bibnamefont {Zhu}}, \ and\
		\bibinfo {author} {\bibfnamefont {L.}~\bibnamefont {He}},\ }\href {\doibase
		10.1103/ktgw-2wx2} {\bibfield  {journal} {\bibinfo  {journal} {Phys. Rev.
				Lett.}\ }\textbf {\bibinfo {volume} {135}},\ \bibinfo {pages} {256401}
		(\bibinfo {year} {2025})}\BibitemShut {NoStop}%
	\bibitem [{\citenamefont {Zou}\ \emph {et~al.}(2019)\citenamefont {Zou},
		\citenamefont {He},\ and\ \citenamefont {Xu}}]{2019MSMNPJ}%
	\BibitemOpen
	\bibfield  {author} {\bibinfo {author} {\bibfnamefont {J.}~\bibnamefont
			{Zou}}, \bibinfo {author} {\bibfnamefont {Z.}~\bibnamefont {He}}, \ and\
		\bibinfo {author} {\bibfnamefont {G.}~\bibnamefont {Xu}},\ }\href {\doibase
		10.1038/s41524-019-0237-5} {\bibfield  {journal} {\bibinfo  {journal} {NPJ
				Comput. Mater.}\ }\textbf {\bibinfo {volume} {5}},\ \bibinfo {pages} {96}
		(\bibinfo {year} {2019})}\BibitemShut {NoStop}%
	\bibitem [{\citenamefont {Zhou}\ \emph {et~al.}(2022)\citenamefont {Zhou},
		\citenamefont {Zhang}, \citenamefont {Yang}, \citenamefont {Li},
		\citenamefont {Feng}, \citenamefont {Mokrousov},\ and\ \citenamefont
		{Yao}}]{2022PRLMSM}%
	\BibitemOpen
	\bibfield  {author} {\bibinfo {author} {\bibfnamefont {X.}~\bibnamefont
			{Zhou}}, \bibinfo {author} {\bibfnamefont {R.-W.}\ \bibnamefont {Zhang}},
		\bibinfo {author} {\bibfnamefont {X.}~\bibnamefont {Yang}}, \bibinfo {author}
		{\bibfnamefont {X.-P.}\ \bibnamefont {Li}}, \bibinfo {author} {\bibfnamefont
			{W.}~\bibnamefont {Feng}}, \bibinfo {author} {\bibfnamefont {Y.}~\bibnamefont
			{Mokrousov}}, \ and\ \bibinfo {author} {\bibfnamefont {Y.}~\bibnamefont
			{Yao}},\ }\href {\doibase 10.1103/PhysRevLett.129.097201} {\bibfield
		{journal} {\bibinfo  {journal} {Phys. Rev. Lett.}\ }\textbf {\bibinfo
			{volume} {129}},\ \bibinfo {pages} {097201} (\bibinfo {year}
		{2022})}\BibitemShut {NoStop}%
	\bibitem [{\citenamefont {Li}\ \emph {et~al.}(2024{\natexlab{a}})\citenamefont
		{Li}, \citenamefont {Huo}, \citenamefont {Wang}, \citenamefont {Jin},
		\citenamefont {Liu}, \citenamefont {Dai}, \citenamefont {Liu}, \citenamefont
		{Zhang},\ and\ \citenamefont {Kou}}]{2024TSMPRA}%
	\BibitemOpen
	\bibfield  {author} {\bibinfo {author} {\bibfnamefont {M.}~\bibnamefont
			{Li}}, \bibinfo {author} {\bibfnamefont {Y.}~\bibnamefont {Huo}}, \bibinfo
		{author} {\bibfnamefont {L.}~\bibnamefont {Wang}}, \bibinfo {author}
		{\bibfnamefont {L.}~\bibnamefont {Jin}}, \bibinfo {author} {\bibfnamefont
			{Y.}~\bibnamefont {Liu}}, \bibinfo {author} {\bibfnamefont {X.}~\bibnamefont
			{Dai}}, \bibinfo {author} {\bibfnamefont {G.}~\bibnamefont {Liu}}, \bibinfo
		{author} {\bibfnamefont {X.}~\bibnamefont {Zhang}}, \ and\ \bibinfo {author}
		{\bibfnamefont {L.}~\bibnamefont {Kou}},\ }\href {\doibase
		10.1103/PhysRevApplied.22.064090} {\bibfield  {journal} {\bibinfo  {journal}
			{Phys. Rev. Appl.}\ }\textbf {\bibinfo {volume} {22}},\ \bibinfo {pages}
		{064090} (\bibinfo {year} {2024}{\natexlab{a}})}\BibitemShut {NoStop}%
	\bibitem [{\citenamefont {Ozawa}\ \emph {et~al.}(2026)\citenamefont {Ozawa},
		\citenamefont {Araki}, \citenamefont {Kobayashi},\ and\ \citenamefont
		{Nomura}}]{2026arxivmsm}%
	\BibitemOpen
	\bibfield  {author} {\bibinfo {author} {\bibfnamefont {A.}~\bibnamefont
			{Ozawa}}, \bibinfo {author} {\bibfnamefont {Y.}~\bibnamefont {Araki}},
		\bibinfo {author} {\bibfnamefont {K.}~\bibnamefont {Kobayashi}}, \ and\
		\bibinfo {author} {\bibfnamefont {K.}~\bibnamefont {Nomura}},\ }\href@noop {}
	{\bibfield  {journal} {\bibinfo  {journal} {arXiv:2603.22568}\ } (\bibinfo
		{year} {2026})}\BibitemShut {NoStop}%
	\bibitem [{\citenamefont {Pupim}\ and\ \citenamefont
		{Scheurer}(2025)}]{2025PRLsuper}%
	\BibitemOpen
	\bibfield  {author} {\bibinfo {author} {\bibfnamefont {L.~V.}\ \bibnamefont
			{Pupim}}\ and\ \bibinfo {author} {\bibfnamefont {M.~S.}\ \bibnamefont
			{Scheurer}},\ }\href {\doibase 10.1103/PhysRevLett.134.146001} {\bibfield
		{journal} {\bibinfo  {journal} {Phys. Rev. Lett.}\ }\textbf {\bibinfo
			{volume} {134}},\ \bibinfo {pages} {146001} (\bibinfo {year}
		{2025})}\BibitemShut {NoStop}%
	\bibitem [{\citenamefont {Song}\ \emph {et~al.}(2025)\citenamefont {Song},
		\citenamefont {Bai}, \citenamefont {Zhou}, \citenamefont {Han}, \citenamefont
		{Reichlova}, \citenamefont {Dil}, \citenamefont {Liu}, \citenamefont {Chen},\
		and\ \citenamefont {Pan}}]{2025NRPAM}%
	\BibitemOpen
	\bibfield  {author} {\bibinfo {author} {\bibfnamefont {C.}~\bibnamefont
			{Song}}, \bibinfo {author} {\bibfnamefont {H.}~\bibnamefont {Bai}}, \bibinfo
		{author} {\bibfnamefont {Z.}~\bibnamefont {Zhou}}, \bibinfo {author}
		{\bibfnamefont {L.}~\bibnamefont {Han}}, \bibinfo {author} {\bibfnamefont
			{H.}~\bibnamefont {Reichlova}}, \bibinfo {author} {\bibfnamefont {J.~H.}\
			\bibnamefont {Dil}}, \bibinfo {author} {\bibfnamefont {J.}~\bibnamefont
			{Liu}}, \bibinfo {author} {\bibfnamefont {X.}~\bibnamefont {Chen}}, \ and\
		\bibinfo {author} {\bibfnamefont {F.}~\bibnamefont {Pan}},\ }\href {\doibase
		10.1038/s41578-025-00779-1} {\bibfield  {journal} {\bibinfo  {journal} {Nat.
				Rev. Mater.}\ }\textbf {\bibinfo {volume} {10}},\ \bibinfo {pages} {473}
		(\bibinfo {year} {2025})}\BibitemShut {NoStop}%
	\bibitem [{\citenamefont {Wang}\ \emph
		{et~al.}(2026{\natexlab{a}})\citenamefont {Wang}, \citenamefont {Yang},
		\citenamefont {Yang}, \citenamefont {Lu}, \citenamefont {Ho}, \citenamefont
		{Wang}, \citenamefont {Ang}, \citenamefont {Cheng},\ and\ \citenamefont
		{Fang}}]{2026AFMAM}%
	\BibitemOpen
	\bibfield  {author} {\bibinfo {author} {\bibfnamefont {J.}~\bibnamefont
			{Wang}}, \bibinfo {author} {\bibfnamefont {X.}~\bibnamefont {Yang}}, \bibinfo
		{author} {\bibfnamefont {Z.}~\bibnamefont {Yang}}, \bibinfo {author}
		{\bibfnamefont {J.}~\bibnamefont {Lu}}, \bibinfo {author} {\bibfnamefont
			{P.}~\bibnamefont {Ho}}, \bibinfo {author} {\bibfnamefont {W.}~\bibnamefont
			{Wang}}, \bibinfo {author} {\bibfnamefont {Y.~S.}\ \bibnamefont {Ang}},
		\bibinfo {author} {\bibfnamefont {Z.}~\bibnamefont {Cheng}}, \ and\ \bibinfo
		{author} {\bibfnamefont {S.}~\bibnamefont {Fang}},\ }\href {\doibase
		https://doi.org/10.1002/adfm.202505145} {\bibfield  {journal} {\bibinfo
			{journal} {Adv. Funct. Mater.}\ }\textbf {\bibinfo {volume} {36}},\ \bibinfo
		{pages} {2505145} (\bibinfo {year} {2026}{\natexlab{a}})}\BibitemShut
	{NoStop}%
	\bibitem [{\citenamefont {Zhang}\ \emph {et~al.}(2026)\citenamefont {Zhang},
		\citenamefont {Cui}, \citenamefont {Wang}, \citenamefont {Duan},
		\citenamefont {Yu},\ and\ \citenamefont {Yao}}]{2026PRBAM}%
	\BibitemOpen
	\bibfield  {author} {\bibinfo {author} {\bibfnamefont {R.-W.}\ \bibnamefont
			{Zhang}}, \bibinfo {author} {\bibfnamefont {C.}~\bibnamefont {Cui}}, \bibinfo
		{author} {\bibfnamefont {Y.}~\bibnamefont {Wang}}, \bibinfo {author}
		{\bibfnamefont {J.}~\bibnamefont {Duan}}, \bibinfo {author} {\bibfnamefont
			{Z.-M.}\ \bibnamefont {Yu}}, \ and\ \bibinfo {author} {\bibfnamefont
			{Y.}~\bibnamefont {Yao}},\ }\href {\doibase 10.1103/s9mm-5662} {\bibfield
		{journal} {\bibinfo  {journal} {Phys. Rev. B}\ }\textbf {\bibinfo {volume}
			{113}},\ \bibinfo {pages} {L161115} (\bibinfo {year} {2026})}\BibitemShut
	{NoStop}%
	\bibitem [{\citenamefont {Huang}\ \emph {et~al.}(2026)\citenamefont {Huang},
		\citenamefont {Qin}, \citenamefont {Zhan}, \citenamefont {Xu}, \citenamefont
		{Ma},\ and\ \citenamefont {Wang}}]{2026PRLAMODD}%
	\BibitemOpen
	\bibfield  {author} {\bibinfo {author} {\bibfnamefont {S.}~\bibnamefont
			{Huang}}, \bibinfo {author} {\bibfnamefont {Z.}~\bibnamefont {Qin}}, \bibinfo
		{author} {\bibfnamefont {F.}~\bibnamefont {Zhan}}, \bibinfo {author}
		{\bibfnamefont {D.-H.}\ \bibnamefont {Xu}}, \bibinfo {author} {\bibfnamefont
			{D.-S.}\ \bibnamefont {Ma}}, \ and\ \bibinfo {author} {\bibfnamefont
			{R.}~\bibnamefont {Wang}},\ }\href {\doibase 10.1103/9346-9jpf} {\bibfield
		{journal} {\bibinfo  {journal} {Phys. Rev. Lett.}\ }\textbf {\bibinfo
			{volume} {136}},\ \bibinfo {pages} {126703} (\bibinfo {year}
		{2026})}\BibitemShut {NoStop}%
	\bibitem [{\citenamefont {Xu}\ \emph {et~al.}(2020)\citenamefont {Xu},
		\citenamefont {Elcoro}, \citenamefont {Song}, \citenamefont {Wieder},
		\citenamefont {Vergniory}, \citenamefont {Regnault}, \citenamefont {Chen},
		\citenamefont {Felser},\ and\ \citenamefont {Bernevig}}]{database_mag}%
	\BibitemOpen
	\bibfield  {author} {\bibinfo {author} {\bibfnamefont {Y.}~\bibnamefont
			{Xu}}, \bibinfo {author} {\bibfnamefont {L.}~\bibnamefont {Elcoro}}, \bibinfo
		{author} {\bibfnamefont {Z.-D.}\ \bibnamefont {Song}}, \bibinfo {author}
		{\bibfnamefont {B.~J.}\ \bibnamefont {Wieder}}, \bibinfo {author}
		{\bibfnamefont {M.}~\bibnamefont {Vergniory}}, \bibinfo {author}
		{\bibfnamefont {N.}~\bibnamefont {Regnault}}, \bibinfo {author}
		{\bibfnamefont {Y.}~\bibnamefont {Chen}}, \bibinfo {author} {\bibfnamefont
			{C.}~\bibnamefont {Felser}}, \ and\ \bibinfo {author} {\bibfnamefont {B.~A.}\
			\bibnamefont {Bernevig}},\ }\href@noop {} {\bibfield  {journal} {\bibinfo
			{journal} {Nature}\ }\textbf {\bibinfo {volume} {586}},\ \bibinfo {pages}
		{702} (\bibinfo {year} {2020})}\BibitemShut {NoStop}%
	\bibitem [{\citenamefont {Cano}\ and\ \citenamefont
		{Bradlyn}(2021)}]{2021ARCTQC}%
	\BibitemOpen
	\bibfield  {author} {\bibinfo {author} {\bibfnamefont {J.}~\bibnamefont
			{Cano}}\ and\ \bibinfo {author} {\bibfnamefont {B.}~\bibnamefont {Bradlyn}},\
	}\href@noop {} {\bibfield  {journal} {\bibinfo  {journal} {Annu. Rev.
				Condens. Matter Phys.}\ }\textbf {\bibinfo {volume} {12}},\ \bibinfo {pages}
		{225} (\bibinfo {year} {2021})}\BibitemShut {NoStop}%
	\bibitem [{\citenamefont {Elcoro}\ \emph {et~al.}(2021)\citenamefont {Elcoro},
		\citenamefont {Wieder}, \citenamefont {Song}, \citenamefont {Xu},
		\citenamefont {Bradlyn},\ and\ \citenamefont {Bernevig}}]{2021NCTQC}%
	\BibitemOpen
	\bibfield  {author} {\bibinfo {author} {\bibfnamefont {L.}~\bibnamefont
			{Elcoro}}, \bibinfo {author} {\bibfnamefont {B.~J.}\ \bibnamefont {Wieder}},
		\bibinfo {author} {\bibfnamefont {Z.}~\bibnamefont {Song}}, \bibinfo {author}
		{\bibfnamefont {Y.}~\bibnamefont {Xu}}, \bibinfo {author} {\bibfnamefont
			{B.}~\bibnamefont {Bradlyn}}, \ and\ \bibinfo {author} {\bibfnamefont
			{B.~A.}\ \bibnamefont {Bernevig}},\ }\href {\doibase
		10.1038/s41467-021-26241-8} {\bibfield  {journal} {\bibinfo  {journal} {Nat.
				Commun.}\ }\textbf {\bibinfo {volume} {12}},\ \bibinfo {pages} {5965}
		(\bibinfo {year} {2021})}\BibitemShut {NoStop}%
	\bibitem [{\citenamefont {Wieder}\ \emph {et~al.}(2022)\citenamefont {Wieder},
		\citenamefont {Bradlyn}, \citenamefont {Cano}, \citenamefont {Wang},
		\citenamefont {Vergniory}, \citenamefont {Elcoro}, \citenamefont {Soluyanov},
		\citenamefont {Felser}, \citenamefont {Neupert}, \citenamefont {Regnault},\
		and\ \citenamefont {Bernevig}}]{2022NRMsymmetry}%
	\BibitemOpen
	\bibfield  {author} {\bibinfo {author} {\bibfnamefont {B.~J.}\ \bibnamefont
			{Wieder}}, \bibinfo {author} {\bibfnamefont {B.}~\bibnamefont {Bradlyn}},
		\bibinfo {author} {\bibfnamefont {J.}~\bibnamefont {Cano}}, \bibinfo {author}
		{\bibfnamefont {Z.}~\bibnamefont {Wang}}, \bibinfo {author} {\bibfnamefont
			{M.~G.}\ \bibnamefont {Vergniory}}, \bibinfo {author} {\bibfnamefont
			{L.}~\bibnamefont {Elcoro}}, \bibinfo {author} {\bibfnamefont {A.~A.}\
			\bibnamefont {Soluyanov}}, \bibinfo {author} {\bibfnamefont {C.}~\bibnamefont
			{Felser}}, \bibinfo {author} {\bibfnamefont {T.}~\bibnamefont {Neupert}},
		\bibinfo {author} {\bibfnamefont {N.}~\bibnamefont {Regnault}}, \ and\
		\bibinfo {author} {\bibfnamefont {B.~A.}\ \bibnamefont {Bernevig}},\ }\href
	{\doibase 10.1038/s41578-021-00380-2} {\bibfield  {journal} {\bibinfo
			{journal} {Nat. Rev. Mater.}\ }\textbf {\bibinfo {volume} {7}},\ \bibinfo
		{pages} {196} (\bibinfo {year} {2022})}\BibitemShut {NoStop}%
	\bibitem [{\citenamefont {Singh}\ \emph {et~al.}(2023)\citenamefont {Singh},
		\citenamefont {Lin},\ and\ \citenamefont {Bansil}}]{2023AMsymmetry}%
	\BibitemOpen
	\bibfield  {author} {\bibinfo {author} {\bibfnamefont {B.}~\bibnamefont
			{Singh}}, \bibinfo {author} {\bibfnamefont {H.}~\bibnamefont {Lin}}, \ and\
		\bibinfo {author} {\bibfnamefont {A.}~\bibnamefont {Bansil}},\ }\href
	{\doibase https://doi.org/10.1002/adma.202201058} {\bibfield  {journal}
		{\bibinfo  {journal} {Adv. Mater.}\ }\textbf {\bibinfo {volume} {35}},\
		\bibinfo {pages} {2201058} (\bibinfo {year} {2023})}\BibitemShut {NoStop}%
	\bibitem [{\citenamefont {Zhou}\ and\ \citenamefont {Zhou}(2024)}]{2024floNL}%
	\BibitemOpen
	\bibfield  {author} {\bibinfo {author} {\bibfnamefont {C.}~\bibnamefont
			{Zhou}}\ and\ \bibinfo {author} {\bibfnamefont {J.}~\bibnamefont {Zhou}},\
	}\href {\doibase 10.1021/acs.nanolett.4c01415} {\bibfield  {journal}
		{\bibinfo  {journal} {Nano Lett.}\ }\textbf {\bibinfo {volume} {24}},\
		\bibinfo {pages} {7311} (\bibinfo {year} {2024})}\BibitemShut {NoStop}%
	\bibitem [{\citenamefont {Dong}\ \emph {et~al.}(2026)\citenamefont {Dong},
		\citenamefont {Xie}, \citenamefont {Bai}, \citenamefont {Wang}, \citenamefont
		{Li}, \citenamefont {Luo}, \citenamefont {Sun}, \citenamefont {Pi},
		\citenamefont {Ma}, \citenamefont {Deng}, \citenamefont {Meng}, \citenamefont
		{Tong}, \citenamefont {Hou}, \citenamefont {Lu}, \citenamefont {Sun},
		\citenamefont {Lu},\ and\ \citenamefont {Feng}}]{2026AFMtopo}%
	\BibitemOpen
	\bibfield  {author} {\bibinfo {author} {\bibfnamefont {S.}~\bibnamefont
			{Dong}}, \bibinfo {author} {\bibfnamefont {C.}~\bibnamefont {Xie}}, \bibinfo
		{author} {\bibfnamefont {Y.}~\bibnamefont {Bai}}, \bibinfo {author}
		{\bibfnamefont {A.}~\bibnamefont {Wang}}, \bibinfo {author} {\bibfnamefont
			{Z.}~\bibnamefont {Li}}, \bibinfo {author} {\bibfnamefont {X.}~\bibnamefont
			{Luo}}, \bibinfo {author} {\bibfnamefont {Y.}~\bibnamefont {Sun}}, \bibinfo
		{author} {\bibfnamefont {L.}~\bibnamefont {Pi}}, \bibinfo {author}
		{\bibfnamefont {M.}~\bibnamefont {Ma}}, \bibinfo {author} {\bibfnamefont
			{Y.}~\bibnamefont {Deng}}, \bibinfo {author} {\bibfnamefont {W.}~\bibnamefont
			{Meng}}, \bibinfo {author} {\bibfnamefont {W.}~\bibnamefont {Tong}}, \bibinfo
		{author} {\bibfnamefont {Y.}~\bibnamefont {Hou}}, \bibinfo {author}
		{\bibfnamefont {Y.}~\bibnamefont {Lu}}, \bibinfo {author} {\bibfnamefont
			{F.-H.}\ \bibnamefont {Sun}}, \bibinfo {author} {\bibfnamefont
			{Q.}~\bibnamefont {Lu}}, \ and\ \bibinfo {author} {\bibfnamefont
			{Q.}~\bibnamefont {Feng}},\ }\href {\doibase
		https://doi.org/10.1002/adfm.74354} {\bibfield  {journal} {\bibinfo
			{journal} {Adv. Funct. Mater.}\ }\textbf {\bibinfo {volume} {36}},\ \bibinfo
		{pages} {e74354} (\bibinfo {year} {2026})}\BibitemShut {NoStop}%
	\bibitem [{\citenamefont {Bai}\ \emph {et~al.}(2025)\citenamefont {Bai},
		\citenamefont {Zhang}, \citenamefont {Feng},\ and\ \citenamefont
		{Yao}}]{2025PRLtype4}%
	\BibitemOpen
	\bibfield  {author} {\bibinfo {author} {\bibfnamefont {L.}~\bibnamefont
			{Bai}}, \bibinfo {author} {\bibfnamefont {R.-W.}\ \bibnamefont {Zhang}},
		\bibinfo {author} {\bibfnamefont {W.}~\bibnamefont {Feng}}, \ and\ \bibinfo
		{author} {\bibfnamefont {Y.}~\bibnamefont {Yao}},\ }\href {\doibase
		10.1103/rn1l-d6cq} {\bibfield  {journal} {\bibinfo  {journal} {Phys. Rev.
				Lett.}\ }\textbf {\bibinfo {volume} {135}},\ \bibinfo {pages} {036702}
		(\bibinfo {year} {2025})}\BibitemShut {NoStop}%
	\bibitem [{\citenamefont {Tian}\ \emph {et~al.}(2026)\citenamefont {Tian},
		\citenamefont {Cui}, \citenamefont {Zhang}, \citenamefont {Duan},
		\citenamefont {Feng},\ and\ \citenamefont {Zhang}}]{2026prltype4}%
	\BibitemOpen
	\bibfield  {author} {\bibinfo {author} {\bibfnamefont {M.}~\bibnamefont
			{Tian}}, \bibinfo {author} {\bibfnamefont {C.}~\bibnamefont {Cui}}, \bibinfo
		{author} {\bibfnamefont {Z.}~\bibnamefont {Zhang}}, \bibinfo {author}
		{\bibfnamefont {J.}~\bibnamefont {Duan}}, \bibinfo {author} {\bibfnamefont
			{W.}~\bibnamefont {Feng}}, \ and\ \bibinfo {author} {\bibfnamefont {R.-W.}\
			\bibnamefont {Zhang}},\ }\href {\doibase 10.1103/jp95-17sz} {\bibfield
		{journal} {\bibinfo  {journal} {Phys. Rev. Lett.}\ }\textbf {\bibinfo
			{volume} {136}},\ \bibinfo {pages} {206701} (\bibinfo {year}
		{2026})}\BibitemShut {NoStop}%
	\bibitem [{\citenamefont {Schindler}\ \emph
		{et~al.}(2018{\natexlab{a}})\citenamefont {Schindler}, \citenamefont {Cook},
		\citenamefont {Vergniory}, \citenamefont {Wang}, \citenamefont {Parkin},
		\citenamefont {Bernevig},\ and\ \citenamefont {Neupert}}]{highorderfirst}%
	\BibitemOpen
	\bibfield  {author} {\bibinfo {author} {\bibfnamefont {F.}~\bibnamefont
			{Schindler}}, \bibinfo {author} {\bibfnamefont {A.~M.}\ \bibnamefont {Cook}},
		\bibinfo {author} {\bibfnamefont {M.~G.}\ \bibnamefont {Vergniory}}, \bibinfo
		{author} {\bibfnamefont {Z.}~\bibnamefont {Wang}}, \bibinfo {author}
		{\bibfnamefont {S.~S.}\ \bibnamefont {Parkin}}, \bibinfo {author}
		{\bibfnamefont {B.~A.}\ \bibnamefont {Bernevig}}, \ and\ \bibinfo {author}
		{\bibfnamefont {T.}~\bibnamefont {Neupert}},\ }\href {\doibase
		10.1126/sciadv.aat0346} {\bibfield  {journal} {\bibinfo  {journal} {Sci.
				Adv.}\ }\textbf {\bibinfo {volume} {4}},\ \bibinfo {pages} {eaat0346}
		(\bibinfo {year} {2018}{\natexlab{a}})}\BibitemShut {NoStop}%
	\bibitem [{\citenamefont {Xie}\ \emph {et~al.}(2021)\citenamefont {Xie},
		\citenamefont {Wang}, \citenamefont {Zhang}, \citenamefont {Zhan},
		\citenamefont {Jiang}, \citenamefont {Lu},\ and\ \citenamefont
		{Chen}}]{2021NRPn1}%
	\BibitemOpen
	\bibfield  {author} {\bibinfo {author} {\bibfnamefont {B.}~\bibnamefont
			{Xie}}, \bibinfo {author} {\bibfnamefont {H.-X.}\ \bibnamefont {Wang}},
		\bibinfo {author} {\bibfnamefont {X.}~\bibnamefont {Zhang}}, \bibinfo
		{author} {\bibfnamefont {P.}~\bibnamefont {Zhan}}, \bibinfo {author}
		{\bibfnamefont {J.-H.}\ \bibnamefont {Jiang}}, \bibinfo {author}
		{\bibfnamefont {M.}~\bibnamefont {Lu}}, \ and\ \bibinfo {author}
		{\bibfnamefont {Y.}~\bibnamefont {Chen}},\ }\href {\doibase
		10.1038/s42254-021-00323-4} {\bibfield  {journal} {\bibinfo  {journal} {Nat.
				Rev. Phys.}\ }\textbf {\bibinfo {volume} {3}},\ \bibinfo {pages} {520}
		(\bibinfo {year} {2021})}\BibitemShut {NoStop}%
	\bibitem [{\citenamefont {Li}\ \emph {et~al.}(2026)\citenamefont {Li},
		\citenamefont {Zhou}, \citenamefont {Zhu}, \citenamefont {Zou}, \citenamefont
		{Cheng},\ and\ \citenamefont {Assouar}}]{2026PRLHOTI}%
	\BibitemOpen
	\bibfield  {author} {\bibinfo {author} {\bibfnamefont {S.-F.}\ \bibnamefont
			{Li}}, \bibinfo {author} {\bibfnamefont {C.-Y.-Y.}\ \bibnamefont {Zhou}},
		\bibinfo {author} {\bibfnamefont {Y.-F.}\ \bibnamefont {Zhu}}, \bibinfo
		{author} {\bibfnamefont {X.-Y.}\ \bibnamefont {Zou}}, \bibinfo {author}
		{\bibfnamefont {J.-C.}\ \bibnamefont {Cheng}}, \ and\ \bibinfo {author}
		{\bibfnamefont {B.}~\bibnamefont {Assouar}},\ }\href {\doibase
		10.1103/74kx-xg78} {\bibfield  {journal} {\bibinfo  {journal} {Phys. Rev.
				Lett.}\ }\textbf {\bibinfo {volume} {136}},\ \bibinfo {pages} {186603}
		(\bibinfo {year} {2026})}\BibitemShut {NoStop}%
	\bibitem [{\citenamefont {Hasan}\ and\ \citenamefont {Kane}(2010)}]{Hasan2010}%
	\BibitemOpen
	\bibfield  {author} {\bibinfo {author} {\bibfnamefont {M.~Z.}\ \bibnamefont
			{Hasan}}\ and\ \bibinfo {author} {\bibfnamefont {C.~L.}\ \bibnamefont
			{Kane}},\ }\href {\doibase 10.1103/RevModPhys.82.3045} {\bibfield  {journal}
		{\bibinfo  {journal} {Rev. Mod. Phys.}\ }\textbf {\bibinfo {volume} {82}},\
		\bibinfo {pages} {3045} (\bibinfo {year} {2010})}\BibitemShut {NoStop}%
	\bibitem [{\citenamefont {Bansil}\ \emph {et~al.}(2016)\citenamefont {Bansil},
		\citenamefont {Lin},\ and\ \citenamefont {Das}}]{2016TIRMP}%
	\BibitemOpen
	\bibfield  {author} {\bibinfo {author} {\bibfnamefont {A.}~\bibnamefont
			{Bansil}}, \bibinfo {author} {\bibfnamefont {H.}~\bibnamefont {Lin}}, \ and\
		\bibinfo {author} {\bibfnamefont {T.}~\bibnamefont {Das}},\ }\href {\doibase
		10.1103/RevModPhys.88.021004} {\bibfield  {journal} {\bibinfo  {journal}
			{Rev. Mod. Phys.}\ }\textbf {\bibinfo {volume} {88}},\ \bibinfo {pages}
		{021004} (\bibinfo {year} {2016})}\BibitemShut {NoStop}%
	\bibitem [{\citenamefont {Liu}(2021)}]{2021cmsreview1}%
	\BibitemOpen
	\bibfield  {author} {\bibinfo {author} {\bibfnamefont {J.}~\bibnamefont
			{Liu}},\ }\href {\doibase https://doi.org/10.1016/j.commatsci.2021.110467}
	{\bibfield  {journal} {\bibinfo  {journal} {Comput. Mater. Sci.}\ }\textbf
		{\bibinfo {volume} {195}},\ \bibinfo {pages} {110467} (\bibinfo {year}
		{2021})}\BibitemShut {NoStop}%
	\bibitem [{\citenamefont {Schindler}\ \emph
		{et~al.}(2018{\natexlab{b}})\citenamefont {Schindler}, \citenamefont {Wang},
		\citenamefont {Vergniory}, \citenamefont {Cook}, \citenamefont {Murani},
		\citenamefont {Sengupta}, \citenamefont {Kasumov}, \citenamefont {Deblock},
		\citenamefont {Jeon}, \citenamefont {Drozdov}, \citenamefont {Bouchiat},
		\citenamefont {Gu{\'{e}}ron}, \citenamefont {Yazdani}, \citenamefont
		{Bernevig},\ and\ \citenamefont {Neupert}}]{Schindler2018}%
	\BibitemOpen
	\bibfield  {author} {\bibinfo {author} {\bibfnamefont {F.}~\bibnamefont
			{Schindler}}, \bibinfo {author} {\bibfnamefont {Z.}~\bibnamefont {Wang}},
		\bibinfo {author} {\bibfnamefont {M.~G.}\ \bibnamefont {Vergniory}}, \bibinfo
		{author} {\bibfnamefont {A.~M.}\ \bibnamefont {Cook}}, \bibinfo {author}
		{\bibfnamefont {A.}~\bibnamefont {Murani}}, \bibinfo {author} {\bibfnamefont
			{S.}~\bibnamefont {Sengupta}}, \bibinfo {author} {\bibfnamefont {A.~Y.}\
			\bibnamefont {Kasumov}}, \bibinfo {author} {\bibfnamefont {R.}~\bibnamefont
			{Deblock}}, \bibinfo {author} {\bibfnamefont {S.}~\bibnamefont {Jeon}},
		\bibinfo {author} {\bibfnamefont {I.}~\bibnamefont {Drozdov}}, \bibinfo
		{author} {\bibfnamefont {H.}~\bibnamefont {Bouchiat}}, \bibinfo {author}
		{\bibfnamefont {S.}~\bibnamefont {Gu{\'{e}}ron}}, \bibinfo {author}
		{\bibfnamefont {A.}~\bibnamefont {Yazdani}}, \bibinfo {author} {\bibfnamefont
			{B.~A.}\ \bibnamefont {Bernevig}}, \ and\ \bibinfo {author} {\bibfnamefont
			{T.}~\bibnamefont {Neupert}},\ }\href {\doibase 10.1038/s41567-018-0224-7}
	{\bibfield  {journal} {\bibinfo  {journal} {Nat. Phys.}\ }\textbf {\bibinfo
			{volume} {14}},\ \bibinfo {pages} {918} (\bibinfo {year}
		{2018}{\natexlab{b}})}\BibitemShut {NoStop}%
	\bibitem [{\citenamefont {Ren}\ \emph {et~al.}(2020)\citenamefont {Ren},
		\citenamefont {Qiao},\ and\ \citenamefont {Niu}}]{highorderrenyafei}%
	\BibitemOpen
	\bibfield  {author} {\bibinfo {author} {\bibfnamefont {Y.}~\bibnamefont
			{Ren}}, \bibinfo {author} {\bibfnamefont {Z.}~\bibnamefont {Qiao}}, \ and\
		\bibinfo {author} {\bibfnamefont {Q.}~\bibnamefont {Niu}},\ }\href {\doibase
		10.1103/PhysRevLett.124.166804} {\bibfield  {journal} {\bibinfo  {journal}
			{Phys. Rev. Lett.}\ }\textbf {\bibinfo {volume} {124}},\ \bibinfo {pages}
		{166804} (\bibinfo {year} {2020})}\BibitemShut {NoStop}%
	\bibitem [{\citenamefont {Hu}\ \emph {et~al.}(2023)\citenamefont {Hu},
		\citenamefont {Zhong}, \citenamefont {Zhang}, \citenamefont {Wang},\ and\
		\citenamefont {Wang}}]{hoti2d}%
	\BibitemOpen
	\bibfield  {author} {\bibinfo {author} {\bibfnamefont {T.}~\bibnamefont
			{Hu}}, \bibinfo {author} {\bibfnamefont {W.}~\bibnamefont {Zhong}}, \bibinfo
		{author} {\bibfnamefont {T.}~\bibnamefont {Zhang}}, \bibinfo {author}
		{\bibfnamefont {W.}~\bibnamefont {Wang}}, \ and\ \bibinfo {author}
		{\bibfnamefont {Z.~F.}\ \bibnamefont {Wang}},\ }\href {\doibase
		10.1038/s41467-023-42884-1} {\bibfield  {journal} {\bibinfo  {journal} {Nat.
				Commun.}\ }\textbf {\bibinfo {volume} {14}},\ \bibinfo {pages} {7092}
		(\bibinfo {year} {2023})}\BibitemShut {NoStop}%
	\bibitem [{\citenamefont {Van~Miert}\ and\ \citenamefont
		{Ortix}(2018)}]{highorderwyckoff}%
	\BibitemOpen
	\bibfield  {author} {\bibinfo {author} {\bibfnamefont {G.}~\bibnamefont
			{Van~Miert}}\ and\ \bibinfo {author} {\bibfnamefont {C.}~\bibnamefont
			{Ortix}},\ }\href@noop {} {\bibfield  {journal} {\bibinfo  {journal} {Phys.
				Rev. B}\ }\textbf {\bibinfo {volume} {98}},\ \bibinfo {pages} {081110}
		(\bibinfo {year} {2018})}\BibitemShut {NoStop}%
	\bibitem [{\citenamefont {Hsu}\ \emph {et~al.}(2019)\citenamefont {Hsu},
		\citenamefont {Zhou}, \citenamefont {Chang}, \citenamefont {Ma},
		\citenamefont {Gedik}, \citenamefont {Bansil}, \citenamefont {Xu},
		\citenamefont {Lin},\ and\ \citenamefont {Fu}}]{Hsu13255}%
	\BibitemOpen
	\bibfield  {author} {\bibinfo {author} {\bibfnamefont {C.-H.}\ \bibnamefont
			{Hsu}}, \bibinfo {author} {\bibfnamefont {X.}~\bibnamefont {Zhou}}, \bibinfo
		{author} {\bibfnamefont {T.-R.}\ \bibnamefont {Chang}}, \bibinfo {author}
		{\bibfnamefont {Q.}~\bibnamefont {Ma}}, \bibinfo {author} {\bibfnamefont
			{N.}~\bibnamefont {Gedik}}, \bibinfo {author} {\bibfnamefont
			{A.}~\bibnamefont {Bansil}}, \bibinfo {author} {\bibfnamefont {S.-Y.}\
			\bibnamefont {Xu}}, \bibinfo {author} {\bibfnamefont {H.}~\bibnamefont
			{Lin}}, \ and\ \bibinfo {author} {\bibfnamefont {L.}~\bibnamefont {Fu}},\
	}\href {\doibase 10.1073/pnas.1900527116} {\bibfield  {journal} {\bibinfo
			{journal} {Proc. Natl. Acad. Sci. U. S. A.}\ }\textbf {\bibinfo {volume}
			{116}},\ \bibinfo {pages} {13255} (\bibinfo {year} {2019})}\BibitemShut
	{NoStop}%
	\bibitem [{\citenamefont {Hu}\ \emph {et~al.}(2024)\citenamefont {Hu},
		\citenamefont {Zhuang},\ and\ \citenamefont {Yang}}]{2024PRLhotisymme}%
	\BibitemOpen
	\bibfield  {author} {\bibinfo {author} {\bibfnamefont {J.}~\bibnamefont
			{Hu}}, \bibinfo {author} {\bibfnamefont {S.}~\bibnamefont {Zhuang}}, \ and\
		\bibinfo {author} {\bibfnamefont {Y.}~\bibnamefont {Yang}},\ }\href {\doibase
		10.1103/PhysRevLett.132.213801} {\bibfield  {journal} {\bibinfo  {journal}
			{Phys. Rev. Lett.}\ }\textbf {\bibinfo {volume} {132}},\ \bibinfo {pages}
		{213801} (\bibinfo {year} {2024})}\BibitemShut {NoStop}%
	\bibitem [{\citenamefont {Chen}\ \emph {et~al.}(2019)\citenamefont {Chen},
		\citenamefont {Deng}, \citenamefont {Shi}, \citenamefont {Zhao},
		\citenamefont {Chen},\ and\ \citenamefont {Dong}}]{2019PRLHOTIPHOTON}%
	\BibitemOpen
	\bibfield  {author} {\bibinfo {author} {\bibfnamefont {X.-D.}\ \bibnamefont
			{Chen}}, \bibinfo {author} {\bibfnamefont {W.-M.}\ \bibnamefont {Deng}},
		\bibinfo {author} {\bibfnamefont {F.-L.}\ \bibnamefont {Shi}}, \bibinfo
		{author} {\bibfnamefont {F.-L.}\ \bibnamefont {Zhao}}, \bibinfo {author}
		{\bibfnamefont {M.}~\bibnamefont {Chen}}, \ and\ \bibinfo {author}
		{\bibfnamefont {J.-W.}\ \bibnamefont {Dong}},\ }\href {\doibase
		10.1103/PhysRevLett.122.233902} {\bibfield  {journal} {\bibinfo  {journal}
			{Phys. Rev. Lett.}\ }\textbf {\bibinfo {volume} {122}},\ \bibinfo {pages}
		{233902} (\bibinfo {year} {2019})}\BibitemShut {NoStop}%
	\bibitem [{\citenamefont {Li}\ \emph {et~al.}(2020{\natexlab{a}})\citenamefont
		{Li}, \citenamefont {Zhirihin}, \citenamefont {Gorlach}, \citenamefont {Ni},
		\citenamefont {Filonov}, \citenamefont {Slobozhanyuk}, \citenamefont {Alù},\
		and\ \citenamefont {Khanikaev}}]{2020NPHOTI}%
	\BibitemOpen
	\bibfield  {author} {\bibinfo {author} {\bibfnamefont {M.}~\bibnamefont
			{Li}}, \bibinfo {author} {\bibfnamefont {D.}~\bibnamefont {Zhirihin}},
		\bibinfo {author} {\bibfnamefont {M.}~\bibnamefont {Gorlach}}, \bibinfo
		{author} {\bibfnamefont {X.}~\bibnamefont {Ni}}, \bibinfo {author}
		{\bibfnamefont {D.}~\bibnamefont {Filonov}}, \bibinfo {author} {\bibfnamefont
			{A.}~\bibnamefont {Slobozhanyuk}}, \bibinfo {author} {\bibfnamefont
			{A.}~\bibnamefont {Alù}}, \ and\ \bibinfo {author} {\bibfnamefont {A.~B.}\
			\bibnamefont {Khanikaev}},\ }\href {\doibase 10.1038/s41566-019-0561-9}
	{\bibfield  {journal} {\bibinfo  {journal} {Nat. Photonics}\ }\textbf
		{\bibinfo {volume} {14}},\ \bibinfo {pages} {89} (\bibinfo {year}
		{2020}{\natexlab{a}})}\BibitemShut {NoStop}%
	\bibitem [{\citenamefont {Wang}\ \emph {et~al.}(2025)\citenamefont {Wang},
		\citenamefont {Meng}, \citenamefont {Yan}, \citenamefont {Zhao},
		\citenamefont {Yang}, \citenamefont {Chen}, \citenamefont {Cheng},
		\citenamefont {Xiao}, \citenamefont {Shum}, \citenamefont {Liu},
		\citenamefont {Yang}, \citenamefont {Chen}, \citenamefont {Xi}, \citenamefont
		{Zhu}, \citenamefont {Xie},\ and\ \citenamefont {Gao}}]{2025NChoti}%
	\BibitemOpen
	\bibfield  {author} {\bibinfo {author} {\bibfnamefont {Z.}~\bibnamefont
			{Wang}}, \bibinfo {author} {\bibfnamefont {Y.}~\bibnamefont {Meng}}, \bibinfo
		{author} {\bibfnamefont {B.}~\bibnamefont {Yan}}, \bibinfo {author}
		{\bibfnamefont {D.}~\bibnamefont {Zhao}}, \bibinfo {author} {\bibfnamefont
			{L.}~\bibnamefont {Yang}}, \bibinfo {author} {\bibfnamefont {J.}~\bibnamefont
			{Chen}}, \bibinfo {author} {\bibfnamefont {M.}~\bibnamefont {Cheng}},
		\bibinfo {author} {\bibfnamefont {T.}~\bibnamefont {Xiao}}, \bibinfo {author}
		{\bibfnamefont {P.~P.}\ \bibnamefont {Shum}}, \bibinfo {author}
		{\bibfnamefont {G.-G.}\ \bibnamefont {Liu}}, \bibinfo {author} {\bibfnamefont
			{Y.}~\bibnamefont {Yang}}, \bibinfo {author} {\bibfnamefont {H.}~\bibnamefont
			{Chen}}, \bibinfo {author} {\bibfnamefont {X.}~\bibnamefont {Xi}}, \bibinfo
		{author} {\bibfnamefont {Z.-X.}\ \bibnamefont {Zhu}}, \bibinfo {author}
		{\bibfnamefont {B.}~\bibnamefont {Xie}}, \ and\ \bibinfo {author}
		{\bibfnamefont {Z.}~\bibnamefont {Gao}},\ }\href {\doibase
		10.1038/s41467-025-58051-7} {\bibfield  {journal} {\bibinfo  {journal} {Nat.
				Commun.}\ }\textbf {\bibinfo {volume} {16}},\ \bibinfo {pages} {3122}
		(\bibinfo {year} {2025})}\BibitemShut {NoStop}%
	\bibitem [{\citenamefont {Xue}\ \emph {et~al.}(2019{\natexlab{a}})\citenamefont
		{Xue}, \citenamefont {Yang}, \citenamefont {Gao}, \citenamefont {Chong},\
		and\ \citenamefont {Zhang}}]{2018NMhotiacou}%
	\BibitemOpen
	\bibfield  {author} {\bibinfo {author} {\bibfnamefont {H.}~\bibnamefont
			{Xue}}, \bibinfo {author} {\bibfnamefont {Y.}~\bibnamefont {Yang}}, \bibinfo
		{author} {\bibfnamefont {F.}~\bibnamefont {Gao}}, \bibinfo {author}
		{\bibfnamefont {Y.}~\bibnamefont {Chong}}, \ and\ \bibinfo {author}
		{\bibfnamefont {B.}~\bibnamefont {Zhang}},\ }\href {\doibase
		10.1038/s41563-018-0251-x} {\bibfield  {journal} {\bibinfo  {journal} {Nat.
				Mater.}\ }\textbf {\bibinfo {volume} {18}},\ \bibinfo {pages} {108} (\bibinfo
		{year} {2019}{\natexlab{a}})}\BibitemShut {NoStop}%
	\bibitem [{\citenamefont {Xue}\ \emph {et~al.}(2019{\natexlab{b}})\citenamefont
		{Xue}, \citenamefont {Yang}, \citenamefont {Liu}, \citenamefont {Gao},
		\citenamefont {Chong},\ and\ \citenamefont {Zhang}}]{2019PRLhotiacou}%
	\BibitemOpen
	\bibfield  {author} {\bibinfo {author} {\bibfnamefont {H.}~\bibnamefont
			{Xue}}, \bibinfo {author} {\bibfnamefont {Y.}~\bibnamefont {Yang}}, \bibinfo
		{author} {\bibfnamefont {G.}~\bibnamefont {Liu}}, \bibinfo {author}
		{\bibfnamefont {F.}~\bibnamefont {Gao}}, \bibinfo {author} {\bibfnamefont
			{Y.}~\bibnamefont {Chong}}, \ and\ \bibinfo {author} {\bibfnamefont
			{B.}~\bibnamefont {Zhang}},\ }\href {\doibase 10.1103/PhysRevLett.122.244301}
	{\bibfield  {journal} {\bibinfo  {journal} {Phys. Rev. Lett.}\ }\textbf
		{\bibinfo {volume} {122}},\ \bibinfo {pages} {244301} (\bibinfo {year}
		{2019}{\natexlab{b}})}\BibitemShut {NoStop}%
	\bibitem [{\citenamefont {Imhof}\ \emph {et~al.}(2018)\citenamefont {Imhof},
		\citenamefont {Berger}, \citenamefont {Bayer}, \citenamefont {Brehm},
		\citenamefont {Molenkamp}, \citenamefont {Kiessling}, \citenamefont
		{Schindler}, \citenamefont {Lee}, \citenamefont {Greiter}, \citenamefont
		{Neupert},\ and\ \citenamefont {Thomale}}]{Imhof2018}%
	\BibitemOpen
	\bibfield  {author} {\bibinfo {author} {\bibfnamefont {S.}~\bibnamefont
			{Imhof}}, \bibinfo {author} {\bibfnamefont {C.}~\bibnamefont {Berger}},
		\bibinfo {author} {\bibfnamefont {F.}~\bibnamefont {Bayer}}, \bibinfo
		{author} {\bibfnamefont {J.}~\bibnamefont {Brehm}}, \bibinfo {author}
		{\bibfnamefont {L.~W.}\ \bibnamefont {Molenkamp}}, \bibinfo {author}
		{\bibfnamefont {T.}~\bibnamefont {Kiessling}}, \bibinfo {author}
		{\bibfnamefont {F.}~\bibnamefont {Schindler}}, \bibinfo {author}
		{\bibfnamefont {C.~H.}\ \bibnamefont {Lee}}, \bibinfo {author} {\bibfnamefont
			{M.}~\bibnamefont {Greiter}}, \bibinfo {author} {\bibfnamefont
			{T.}~\bibnamefont {Neupert}}, \ and\ \bibinfo {author} {\bibfnamefont
			{R.}~\bibnamefont {Thomale}},\ }\href
	{http://dx.doi.org/10.1038/s41567-018-0246-1} {\bibfield  {journal} {\bibinfo
			{journal} {Nat. Phys.}\ }\textbf {\bibinfo {volume} {14}},\ \bibinfo {pages}
		{925} (\bibinfo {year} {2018})}\BibitemShut {NoStop}%
	\bibitem [{\citenamefont {Peterson}\ \emph {et~al.}(2018)\citenamefont
		{Peterson}, \citenamefont {Benalcazar}, \citenamefont {Hughes},\ and\
		\citenamefont {Bahl}}]{Peterson2018}%
	\BibitemOpen
	\bibfield  {author} {\bibinfo {author} {\bibfnamefont {C.~W.}\ \bibnamefont
			{Peterson}}, \bibinfo {author} {\bibfnamefont {W.~A.}\ \bibnamefont
			{Benalcazar}}, \bibinfo {author} {\bibfnamefont {T.~L.}\ \bibnamefont
			{Hughes}}, \ and\ \bibinfo {author} {\bibfnamefont {G.}~\bibnamefont
			{Bahl}},\ }\href@noop {} {\bibfield  {journal} {\bibinfo  {journal} {Nature}\
		}\textbf {\bibinfo {volume} {555}},\ \bibinfo {pages} {346} (\bibinfo {year}
		{2018})}\BibitemShut {NoStop}%
	\bibitem [{\citenamefont {Li}\ \emph {et~al.}(2023)\citenamefont {Li},
		\citenamefont {Zhang}, \citenamefont {Mei}, \citenamefont {Xie},
		\citenamefont {Lu}, \citenamefont {Ma}, \citenamefont {Xiao},\ and\
		\citenamefont {Jia}}]{2023PRAHOTIcir}%
	\BibitemOpen
	\bibfield  {author} {\bibinfo {author} {\bibfnamefont {Y.}~\bibnamefont
			{Li}}, \bibinfo {author} {\bibfnamefont {J.-H.}\ \bibnamefont {Zhang}},
		\bibinfo {author} {\bibfnamefont {F.}~\bibnamefont {Mei}}, \bibinfo {author}
		{\bibfnamefont {B.}~\bibnamefont {Xie}}, \bibinfo {author} {\bibfnamefont
			{M.-H.}\ \bibnamefont {Lu}}, \bibinfo {author} {\bibfnamefont
			{J.}~\bibnamefont {Ma}}, \bibinfo {author} {\bibfnamefont {L.}~\bibnamefont
			{Xiao}}, \ and\ \bibinfo {author} {\bibfnamefont {S.}~\bibnamefont {Jia}},\
	}\href {\doibase 10.1103/PhysRevApplied.20.064042} {\bibfield  {journal}
		{\bibinfo  {journal} {Phys. Rev. Appl.}\ }\textbf {\bibinfo {volume} {20}},\
		\bibinfo {pages} {064042} (\bibinfo {year} {2023})}\BibitemShut {NoStop}%
	\bibitem [{\citenamefont {Xu}\ \emph {et~al.}(2019)\citenamefont {Xu},
		\citenamefont {Song}, \citenamefont {Wang}, \citenamefont {Weng},\ and\
		\citenamefont {Dai}}]{higherordereuln2as2}%
	\BibitemOpen
	\bibfield  {author} {\bibinfo {author} {\bibfnamefont {Y.}~\bibnamefont
			{Xu}}, \bibinfo {author} {\bibfnamefont {Z.}~\bibnamefont {Song}}, \bibinfo
		{author} {\bibfnamefont {Z.}~\bibnamefont {Wang}}, \bibinfo {author}
		{\bibfnamefont {H.}~\bibnamefont {Weng}}, \ and\ \bibinfo {author}
		{\bibfnamefont {X.}~\bibnamefont {Dai}},\ }\href {\doibase
		10.1103/PhysRevLett.122.256402} {\bibfield  {journal} {\bibinfo  {journal}
			{Phys. Rev. Lett.}\ }\textbf {\bibinfo {volume} {122}},\ \bibinfo {pages}
		{256402} (\bibinfo {year} {2019})}\BibitemShut {NoStop}%
	\bibitem [{\citenamefont {Luo}\ \emph {et~al.}(2022)\citenamefont {Luo},
		\citenamefont {Song},\ and\ \citenamefont {Xu}}]{2022npjfese}%
	\BibitemOpen
	\bibfield  {author} {\bibinfo {author} {\bibfnamefont {A.}~\bibnamefont
			{Luo}}, \bibinfo {author} {\bibfnamefont {Z.}~\bibnamefont {Song}}, \ and\
		\bibinfo {author} {\bibfnamefont {G.}~\bibnamefont {Xu}},\ }\href {\doibase
		10.1038/s41524-022-00707-9} {\bibfield  {journal} {\bibinfo  {journal} {NPJ
				Comput. Mater.}\ }\textbf {\bibinfo {volume} {8}},\ \bibinfo {pages} {26}
		(\bibinfo {year} {2022})}\BibitemShut {NoStop}%
	\bibitem [{\citenamefont {Chen}\ \emph {et~al.}(2024)\citenamefont {Chen},
		\citenamefont {Li}, \citenamefont {Bai}, \citenamefont {Mao}, \citenamefont
		{Zeer}, \citenamefont {Go}, \citenamefont {Dai}, \citenamefont {Huang},
		\citenamefont {Mokrousov},\ and\ \citenamefont {Niu}}]{2024NLohe}%
	\BibitemOpen
	\bibfield  {author} {\bibinfo {author} {\bibfnamefont {Z.}~\bibnamefont
			{Chen}}, \bibinfo {author} {\bibfnamefont {R.}~\bibnamefont {Li}}, \bibinfo
		{author} {\bibfnamefont {Y.}~\bibnamefont {Bai}}, \bibinfo {author}
		{\bibfnamefont {N.}~\bibnamefont {Mao}}, \bibinfo {author} {\bibfnamefont
			{M.}~\bibnamefont {Zeer}}, \bibinfo {author} {\bibfnamefont {D.}~\bibnamefont
			{Go}}, \bibinfo {author} {\bibfnamefont {Y.}~\bibnamefont {Dai}}, \bibinfo
		{author} {\bibfnamefont {B.}~\bibnamefont {Huang}}, \bibinfo {author}
		{\bibfnamefont {Y.}~\bibnamefont {Mokrousov}}, \ and\ \bibinfo {author}
		{\bibfnamefont {C.}~\bibnamefont {Niu}},\ }\href {\doibase
		10.1021/acs.nanolett.3c05129} {\bibfield  {journal} {\bibinfo  {journal}
			{Nano Lett.}\ }\textbf {\bibinfo {volume} {24}},\ \bibinfo {pages} {4826}
		(\bibinfo {year} {2024})}\BibitemShut {NoStop}%
	\bibitem [{\citenamefont {Li}\ \emph {et~al.}(2024{\natexlab{b}})\citenamefont
		{Li}, \citenamefont {Liu},\ and\ \citenamefont {Liu}}]{2024PRBAMHOTI}%
	\BibitemOpen
	\bibfield  {author} {\bibinfo {author} {\bibfnamefont {Y.-X.}\ \bibnamefont
			{Li}}, \bibinfo {author} {\bibfnamefont {Y.}~\bibnamefont {Liu}}, \ and\
		\bibinfo {author} {\bibfnamefont {C.-C.}\ \bibnamefont {Liu}},\ }\href
	{\doibase 10.1103/PhysRevB.109.L201109} {\bibfield  {journal} {\bibinfo
			{journal} {Phys. Rev. B}\ }\textbf {\bibinfo {volume} {109}},\ \bibinfo
		{pages} {L201109} (\bibinfo {year} {2024}{\natexlab{b}})}\BibitemShut
	{NoStop}%
	\bibitem [{\citenamefont {Wang}\ \emph
		{et~al.}(2026{\natexlab{b}})\citenamefont {Wang}, \citenamefont {Ghosh},
		\citenamefont {Tao}, \citenamefont {Ma},\ and\ \citenamefont
		{Song}}]{2026ASAM}%
	\BibitemOpen
	\bibfield  {author} {\bibinfo {author} {\bibfnamefont {D.}~\bibnamefont
			{Wang}}, \bibinfo {author} {\bibfnamefont {A.~K.}\ \bibnamefont {Ghosh}},
		\bibinfo {author} {\bibfnamefont {Y.}~\bibnamefont {Tao}}, \bibinfo {author}
		{\bibfnamefont {F.}~\bibnamefont {Ma}}, \ and\ \bibinfo {author}
		{\bibfnamefont {C.}~\bibnamefont {Song}},\ }\href {\doibase
		https://doi.org/10.1002/advs.202522203} {\bibfield  {journal} {\bibinfo
			{journal} {Adv. Sci.}\ }\textbf {\bibinfo {volume} {13}},\ \bibinfo {pages}
		{e22203} (\bibinfo {year} {2026}{\natexlab{b}})}\BibitemShut {NoStop}%
	\bibitem [{\citenamefont {Kresse}\ and\ \citenamefont {Hafner}(1993)}]{Kresse}%
	\BibitemOpen
	\bibfield  {author} {\bibinfo {author} {\bibfnamefont {G.}~\bibnamefont
			{Kresse}}\ and\ \bibinfo {author} {\bibfnamefont {J.}~\bibnamefont
			{Hafner}},\ }\href {\doibase 10.1103/PhysRevB.47.558} {\bibfield  {journal}
		{\bibinfo  {journal} {Phys. Rev. B}\ }\textbf {\bibinfo {volume} {47}},\
		\bibinfo {pages} {558} (\bibinfo {year} {1993})}\BibitemShut {NoStop}%
	\bibitem [{\citenamefont {Kresse}\ and\ \citenamefont
		{Furthmüller}(1996)}]{Kresse1}%
	\BibitemOpen
	\bibfield  {author} {\bibinfo {author} {\bibfnamefont {G.}~\bibnamefont
			{Kresse}}\ and\ \bibinfo {author} {\bibfnamefont {J.}~\bibnamefont
			{Furthmüller}},\ }\href@noop {} {\bibfield  {journal} {\bibinfo  {journal}
			{Phys. Rev. B}\ }\textbf {\bibinfo {volume} {54}},\ \bibinfo {pages} {11169}
		(\bibinfo {year} {1996})}\BibitemShut {NoStop}%
	\bibitem [{\citenamefont {Perdew}\ \emph {et~al.}(1996)\citenamefont {Perdew},
		\citenamefont {Burke},\ and\ \citenamefont {Ernzerhof}}]{pbevasp}%
	\BibitemOpen
	\bibfield  {author} {\bibinfo {author} {\bibfnamefont {J.~P.}\ \bibnamefont
			{Perdew}}, \bibinfo {author} {\bibfnamefont {K.}~\bibnamefont {Burke}}, \
		and\ \bibinfo {author} {\bibfnamefont {M.}~\bibnamefont {Ernzerhof}},\
	}\href@noop {} {\bibfield  {journal} {\bibinfo  {journal} {Phys. Rev. Lett.}\
		}\textbf {\bibinfo {volume} {77}},\ \bibinfo {pages} {3865} (\bibinfo {year}
		{1996})}\BibitemShut {NoStop}%
	\bibitem [{\citenamefont {Pizzi}\ \emph {et~al.}(2020)\citenamefont {Pizzi},
		\citenamefont {Vitale}, \citenamefont {Arita}, \citenamefont {Blügel},
		\citenamefont {Freimuth}, \citenamefont {Géranton}, \citenamefont
		{Gibertini}, \citenamefont {Gresch}, \citenamefont {Johnson}, \citenamefont
		{Koretsune}, \citenamefont {Ibañez-Azpiroz}, \citenamefont {Lee},
		\citenamefont {Lihm}, \citenamefont {Marchand}, \citenamefont {Marrazzo},
		\citenamefont {Mokrousov}, \citenamefont {Mustafa}, \citenamefont {Nohara},
		\citenamefont {Nomura}, \citenamefont {Paulatto}, \citenamefont {Poncé},
		\citenamefont {Ponweiser}, \citenamefont {Qiao}, \citenamefont {Thöle},
		\citenamefont {Tsirkin}, \citenamefont {Wierzbowska}, \citenamefont
		{Marzari}, \citenamefont {Vanderbilt}, \citenamefont {Souza}, \citenamefont
		{Mostofi},\ and\ \citenamefont {Yates}}]{wannier90}%
	\BibitemOpen
	\bibfield  {author} {\bibinfo {author} {\bibfnamefont {G.}~\bibnamefont
			{Pizzi}}, \bibinfo {author} {\bibfnamefont {V.}~\bibnamefont {Vitale}},
		\bibinfo {author} {\bibfnamefont {R.}~\bibnamefont {Arita}}, \bibinfo
		{author} {\bibfnamefont {S.}~\bibnamefont {Blügel}}, \bibinfo {author}
		{\bibfnamefont {F.}~\bibnamefont {Freimuth}}, \bibinfo {author}
		{\bibfnamefont {G.}~\bibnamefont {Géranton}}, \bibinfo {author}
		{\bibfnamefont {M.}~\bibnamefont {Gibertini}}, \bibinfo {author}
		{\bibfnamefont {D.}~\bibnamefont {Gresch}}, \bibinfo {author} {\bibfnamefont
			{C.}~\bibnamefont {Johnson}}, \bibinfo {author} {\bibfnamefont
			{T.}~\bibnamefont {Koretsune}}, \bibinfo {author} {\bibfnamefont
			{J.}~\bibnamefont {Ibañez-Azpiroz}}, \bibinfo {author} {\bibfnamefont
			{H.}~\bibnamefont {Lee}}, \bibinfo {author} {\bibfnamefont {J.-M.}\
			\bibnamefont {Lihm}}, \bibinfo {author} {\bibfnamefont {D.}~\bibnamefont
			{Marchand}}, \bibinfo {author} {\bibfnamefont {A.}~\bibnamefont {Marrazzo}},
		\bibinfo {author} {\bibfnamefont {Y.}~\bibnamefont {Mokrousov}}, \bibinfo
		{author} {\bibfnamefont {J.~I.}\ \bibnamefont {Mustafa}}, \bibinfo {author}
		{\bibfnamefont {Y.}~\bibnamefont {Nohara}}, \bibinfo {author} {\bibfnamefont
			{Y.}~\bibnamefont {Nomura}}, \bibinfo {author} {\bibfnamefont
			{L.}~\bibnamefont {Paulatto}}, \bibinfo {author} {\bibfnamefont
			{S.}~\bibnamefont {Poncé}}, \bibinfo {author} {\bibfnamefont
			{T.}~\bibnamefont {Ponweiser}}, \bibinfo {author} {\bibfnamefont
			{J.}~\bibnamefont {Qiao}}, \bibinfo {author} {\bibfnamefont {F.}~\bibnamefont
			{Thöle}}, \bibinfo {author} {\bibfnamefont {S.~S.}\ \bibnamefont {Tsirkin}},
		\bibinfo {author} {\bibfnamefont {M.}~\bibnamefont {Wierzbowska}}, \bibinfo
		{author} {\bibfnamefont {N.}~\bibnamefont {Marzari}}, \bibinfo {author}
		{\bibfnamefont {D.}~\bibnamefont {Vanderbilt}}, \bibinfo {author}
		{\bibfnamefont {I.}~\bibnamefont {Souza}}, \bibinfo {author} {\bibfnamefont
			{A.~A.}\ \bibnamefont {Mostofi}}, \ and\ \bibinfo {author} {\bibfnamefont
			{J.~R.}\ \bibnamefont {Yates}},\ }\href {\doibase 10.1088/1361-648x/ab51ff}
	{\bibfield  {journal} {\bibinfo  {journal} {J. Phys.: Condens. Matter.}\
		}\textbf {\bibinfo {volume} {32}},\ \bibinfo {pages} {165902} (\bibinfo
		{year} {2020})}\BibitemShut {NoStop}%
	\bibitem [{\citenamefont {Wu}\ \emph {et~al.}(2018)\citenamefont {Wu},
		\citenamefont {Zhang}, \citenamefont {Song}, \citenamefont {Troyer},\ and\
		\citenamefont {Soluyanov}}]{WU2017}%
	\BibitemOpen
	\bibfield  {author} {\bibinfo {author} {\bibfnamefont {Q.}~\bibnamefont
			{Wu}}, \bibinfo {author} {\bibfnamefont {S.}~\bibnamefont {Zhang}}, \bibinfo
		{author} {\bibfnamefont {H.-F.}\ \bibnamefont {Song}}, \bibinfo {author}
		{\bibfnamefont {M.}~\bibnamefont {Troyer}}, \ and\ \bibinfo {author}
		{\bibfnamefont {A.~A.}\ \bibnamefont {Soluyanov}},\ }\href {\doibase
		https://doi.org/10.1016/j.cpc.2017.09.033} {\bibfield  {journal} {\bibinfo
			{journal} {Comput. Phys. Commun.}\ }\textbf {\bibinfo {volume} {224}},\
		\bibinfo {pages} {405 } (\bibinfo {year} {2018})}\BibitemShut {NoStop}%
	\bibitem [{\citenamefont {Benalcazar}\ \emph {et~al.}(2019)\citenamefont
		{Benalcazar}, \citenamefont {Li},\ and\ \citenamefont
		{Hughes}}]{highorderinvariants}%
	\BibitemOpen
	\bibfield  {author} {\bibinfo {author} {\bibfnamefont {W.~A.}\ \bibnamefont
			{Benalcazar}}, \bibinfo {author} {\bibfnamefont {T.}~\bibnamefont {Li}}, \
		and\ \bibinfo {author} {\bibfnamefont {T.~L.}\ \bibnamefont {Hughes}},\
	}\href@noop {} {\bibfield  {journal} {\bibinfo  {journal} {Phys. Rev. B}\
		}\textbf {\bibinfo {volume} {99}},\ \bibinfo {pages} {245151} (\bibinfo
		{year} {2019})}\BibitemShut {NoStop}%
	\bibitem [{\citenamefont {Schindler}\ \emph {et~al.}(2019)\citenamefont
		{Schindler}, \citenamefont {Brzezińska}, \citenamefont {Benalcazar},
		\citenamefont {Iraola}, \citenamefont {Bouhon}, \citenamefont {Tsirkin},
		\citenamefont {Vergniory},\ and\ \citenamefont {Neupert}}]{c3invariant}%
	\BibitemOpen
	\bibfield  {author} {\bibinfo {author} {\bibfnamefont {F.}~\bibnamefont
			{Schindler}}, \bibinfo {author} {\bibfnamefont {M.}~\bibnamefont
			{Brzezińska}}, \bibinfo {author} {\bibfnamefont {W.~A.}\ \bibnamefont
			{Benalcazar}}, \bibinfo {author} {\bibfnamefont {M.}~\bibnamefont {Iraola}},
		\bibinfo {author} {\bibfnamefont {A.}~\bibnamefont {Bouhon}}, \bibinfo
		{author} {\bibfnamefont {S.~S.}\ \bibnamefont {Tsirkin}}, \bibinfo {author}
		{\bibfnamefont {M.~G.}\ \bibnamefont {Vergniory}}, \ and\ \bibinfo {author}
		{\bibfnamefont {T.}~\bibnamefont {Neupert}},\ }\href {\doibase
		10.1103/PhysRevResearch.1.033074} {\bibfield  {journal} {\bibinfo  {journal}
			{Phys. Rev. Res.}\ }\textbf {\bibinfo {volume} {1}},\ \bibinfo {pages}
		{033074} (\bibinfo {year} {2019})}\BibitemShut {NoStop}%
	\bibitem [{\citenamefont {Li}\ \emph {et~al.}(2020{\natexlab{b}})\citenamefont
		{Li}, \citenamefont {Zhu}, \citenamefont {Benalcazar},\ and\ \citenamefont
		{Hughes}}]{higher2020invariant}%
	\BibitemOpen
	\bibfield  {author} {\bibinfo {author} {\bibfnamefont {T.}~\bibnamefont
			{Li}}, \bibinfo {author} {\bibfnamefont {P.}~\bibnamefont {Zhu}}, \bibinfo
		{author} {\bibfnamefont {W.~A.}\ \bibnamefont {Benalcazar}}, \ and\ \bibinfo
		{author} {\bibfnamefont {T.~L.}\ \bibnamefont {Hughes}},\ }\href {\doibase
		10.1103/PhysRevB.101.115115} {\bibfield  {journal} {\bibinfo  {journal}
			{Phys. Rev. B}\ }\textbf {\bibinfo {volume} {101}},\ \bibinfo {pages}
		{115115} (\bibinfo {year} {2020}{\natexlab{b}})}\BibitemShut {NoStop}%
	\bibitem [{\citenamefont {Qian}\ \emph {et~al.}(2022)\citenamefont {Qian},
		\citenamefont {Liu}, \citenamefont {Liu},\ and\ \citenamefont
		{Yao}}]{2022PRBHOTI1}%
	\BibitemOpen
	\bibfield  {author} {\bibinfo {author} {\bibfnamefont {S.}~\bibnamefont
			{Qian}}, \bibinfo {author} {\bibfnamefont {G.-B.}\ \bibnamefont {Liu}},
		\bibinfo {author} {\bibfnamefont {C.-C.}\ \bibnamefont {Liu}}, \ and\
		\bibinfo {author} {\bibfnamefont {Y.}~\bibnamefont {Yao}},\ }\href {\doibase
		10.1103/PhysRevB.105.045417} {\bibfield  {journal} {\bibinfo  {journal}
			{Phys. Rev. B}\ }\textbf {\bibinfo {volume} {105}},\ \bibinfo {pages}
		{045417} (\bibinfo {year} {2022})}\BibitemShut {NoStop}%
\end{thebibliography}

%

\end{document}